# A nanolaser with extreme dielectric confinement

Meng Xiong[1,2]†, Yi Yu[1,2]*†, Yury Berdnikov[1], Simon Klinck Borregaard[1,2], Adrian Holm Dubré[1], Elizaveta Semenova[1,2], Kresten Yvind[1,2], and Jesper Mørk[1,2]*

[1]Department of Electrical and Photonics Engineering, Technical University of Denmark, Ørsteds Plads 345A, DK-2800 Lyngby, Denmark.

[2]NanoPhoton - Center for Nanophotonics, Ørsteds Plads 345A, DK-2800 Lyngby, Denmark.

†These authors contribute equally to this work.

*Corresponding authors: yiyu@dtu.dk; jesm@dtu.dk

The interaction between light and matter can be enhanced by spatially concentrating the light field to boost the photon energy density and increasing the photon dwell time to prolong energy transfer between light and matter. Traditionally, strong spatial light localization has been achieved using plasmonics [1], which, despite its effectiveness [2], [3], entails ohmic losses [4]. Recent advances in nanostructured dielectrics [5]-[13] offer an avenue for achieving strong light confinement without metallic losses. However, previous studies primarily focused on minimizing the optical mode volume without adequately addressing light-matter interactions. Here, we develop a nanolaser that simultaneously localizes the electromagnetic field and excited carriers within the same region of a dielectric nanobridge. This extreme dielectric confinement of both light and matter achieves a mode volume below the diffraction limit and a subwavelength carrier volume without the introduction of lateral quantum confinement, enabling continuous-wave lasing at room-temperature. Moreover, we observe a strong correlation between the mode field and carrier distribution, and unexpectedly, the enhanced mode field localization automatically leads to more pronounced carrier localization, promoting self-alignment of light and matter, which significantly reduces the laser threshold. We quantify the intensified light-matter interaction with a newly proposed interaction volume, which generalizes the concept of mode volume to a broad class of active media. Our work lays the ground for developing ultra-efficient optoelectronic devices by greatly enhancing light-matter interactions through advanced material nanostructuring.

**Main**

The capture of light in space and time for enhanced light-matter interactions underpins fundamental quantum optics and a myriad of applications, particularly lasers, which facilitate global connectivity by converting electrical signals into optical ones for long-distance communication. Present lasers, by electronic standards, are large and energy-demanding, impeding their application for on-chip optical links to minimize losses from electrical interconnects [14]-[18]. Such losses, exacerbated by the rise of AI [19], escalate the global energy consumption [20]. Therefore, developing ultra-efficient microscopic lasers is imperative, which necessitates small cavities capable of strong spatial light compression while maintaining low losses.

Since E. M. Purcell's insight that an atom's emission rate can be dramatically enhanced in a small cavity concentrating the electromagnetic vacuum field [21], minimizing the optical mode volume ($V_{mod}$) has been an important research direction. Metallic structures supporting plasmonic excitations have realized mode volumes below the so-called diffraction limit [2], quantified by $(\lambda/(2n))^3$, where $\lambda$ and $n$ are the operating wavelength and refractive index. However, plasmonics suffer from ohmic losses resulting in low quality-factors ($Q$-factors) [4]. Conversely, a new class of dielectric nanostructures featuring sharp-edged geometries can confine light into a sub-diffraction-limit "hotspot" through electromagnetic boundary conditions while preserving high $Q$-factors [5]-[11]. Recent implementations in passive nanobeams and photonic-crystal (PhC) lasers have reached exceptionally small $V_{mod}$ [12], [13], but their mode fields are maximized in air, mirroring void structures [22]-[28]. Thus, the light-matter interaction is compromised, necessitating, e.g., pulsed pumping to achieve lasing [13], [26], [28].

In this work, we demonstrate a nanocavity with extreme dielectric confinement (EDC) of both photons and electrons. We find that the mechanism used to spatially localize the optical field also concentrates the distribution of excited electrons. This co-localization, coupled with high gain and cavity $Q$-factor, significantly enhances the interaction between light and matter. It enables the realization of a nanolaser with a mode volume below the diffraction limit and an ultra-low threshold, operating at room-temperature under continuous-wave (CW) pumping.

**Interaction volume**

While reducing $V_{mod}$ is key for improving the efficiency of quantum light sources with a quantum emitter positioned at the antinode of the cavity mode field [29], [30], this metric inadequately quantifies light-matter interactions in optoelectronic devices like lasers that typically contain many emitters distributed over a region larger than the mode volume. Merely reducing $V_{mod}$ does not enhance light-matter interactions unless the "matter", represented by carriers (electrons and holes), is also spatially localized. This necessity stems from the dependence of light-matter interaction strength on their spatial overlap

$$\frac{1}{V_{\mathbf{I}}} = \int N_p(\mathbf{r}) N_c(\mathbf{r}) d\mathbf{r}, \tag{1}$$

where $N_p(\mathbf{r})$ and $N_c(\mathbf{r})$ are the normalized photon and carrier densities. Here, we introduce the interaction volume ($V_{\mathbf{I}}$), a metric quantifying this overlap and reflecting the degree of spatial confinement for both photons and carriers. For a laser, the threshold power ($P_{th}$) is proportional to the threshold carrier number ($n_{c,th}$), which is proportional to $V_{\mathbf{I}}$ (SI, section D):

$$P_{th} \propto n_{c,th}, \quad \text{with} \quad n_{c,th} = V_{\mathbf{I}} \frac{\omega_c}{gQ} + n_{tr}. \tag{2}$$

Here, $n_{tr}$ is the number of carriers needed to reach transparency, where stimulated absorption and emission exactly balance, and $V_{\mathbf{I}}\omega_c/(gQ)$ is the additional number of excited carriers required to offset the cavity losses, thereby achieving lasing. The parameters $\omega_c$, $g$, and $Q$ are the laser frequency, differential gain coefficient, and cavity $Q$-factor, respectively. Eq. (2) shows that the minimum laser threshold should target a minimized $V_{\mathbf{I}}$ rather than $V_{mod}$, while simultaneously maximizing $Q$ and $g$. In conventional lasers where the carrier density is relatively uniform across the active region, $N_c(\mathbf{r})$ can be factored out of the volume integral of Eq. (1). This reduces $V_{\mathbf{I}}$ to the optical volume used in traditional laser theories [31], [32] (SI, section D.1), which is inversely proportional to the optical confinement factor ($\Gamma$). In the opposite limit of a dimensionless point-dipole, $V_{\mathbf{I}}$ approaches $V_{mod}$ (SI, section D.2). In this scenario, an ultrasmall active material [33]-[35] must be precisely positioned at the "hotspot". Besides the difficulty of such nanoscale alignment, the imposed lateral quantum confinement restricts the total carrier number available ($n_{c,a}$) for stimulated emission ($n_{c,a}$ must be larger than $n_{c,th}$), thereby limiting the available optical gain, which reduces the output power and possibly impedes lasing. Similarly, void structures, which can house quantum emitters inside the voids to provide gain, face the same challenges. In contrast, the EDC geometry with an extended active medium offers high gain without alignment issues. Furthermore, as we will show, $N_c(\mathbf{r})$ in the EDC cavity can vary considerably even without lateral quantum confinement, leading to a substantially reduced $V_I$ while preserving a high gain and $Q$, thereby significantly reducing $P_{th}$.

**Devices and lasing characteristics**

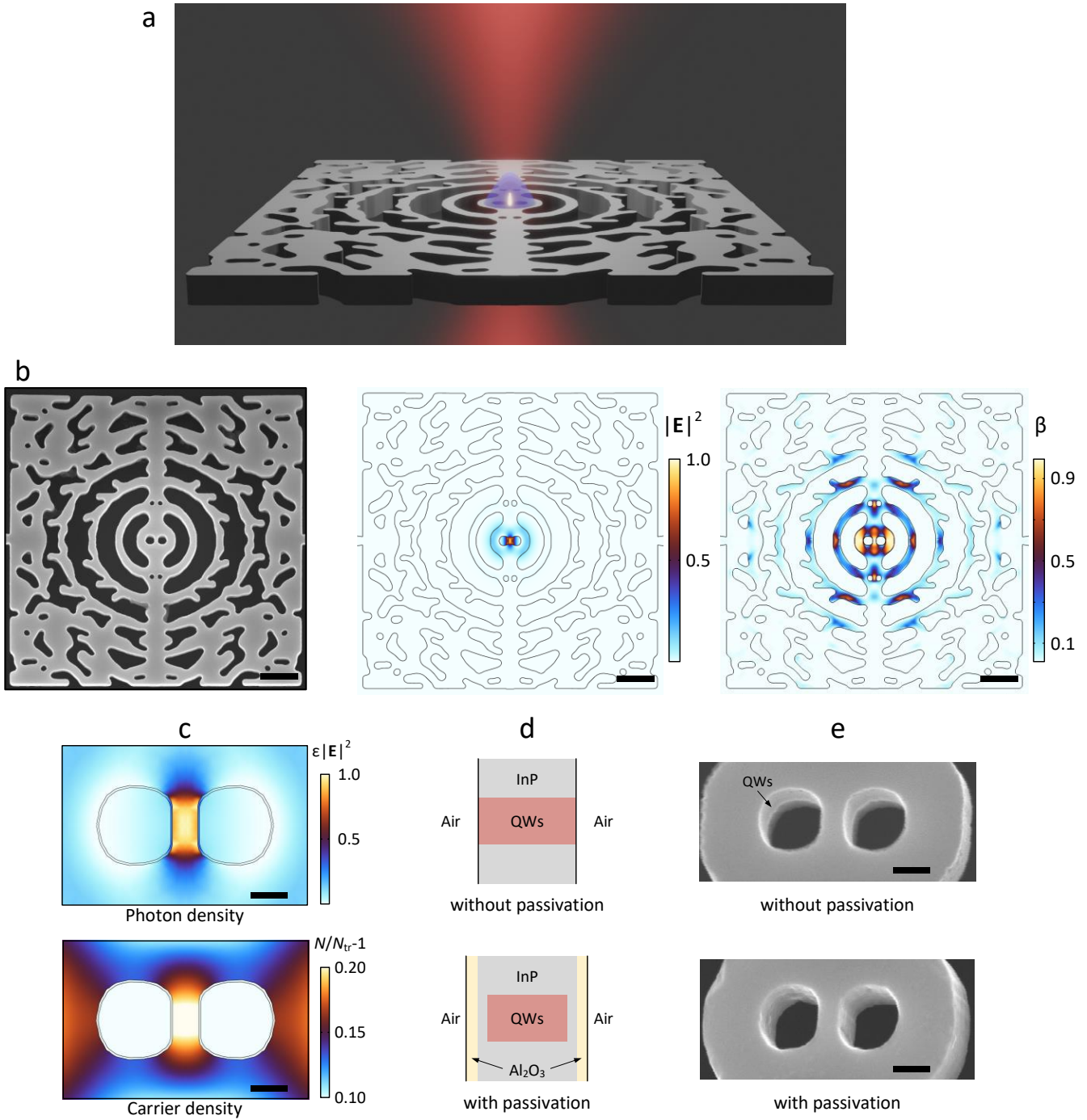

**Fig. 1: Nanolaser based on extreme dielectric confinement (EDC). a.** Schematic of an EDC laser under optical pumping with a focused Gaussian beam (red light), generating a concentrated carrier region (blue shading) with enhanced spatial overlap with photons (bright dot). **b.** Left: Scanning electron microscope (SEM) image of a fabricated EDC laser based on an InP membrane embedded with quantum-wells (QWs). Middle: Simulated lasing mode of the EDC cavity, showing that the electrical field ($|\mathbf{E}|^2$) is strongly localized at a "hotspot" within a central nanobridge, leading to an optical mode volume below the diffraction limit. Right: Corresponding calculated

spontaneous emission factor ($\beta$), illustrating a high value approaching unity at the cavity centre, attributed to the small mode volume. The enhanced $|\mathbf{E}|^2$ and $\beta$ suggest improved light-matter interactions when carriers are also concentrated at the cavity centre. Scale bar: 1 μm. **c.** Calculated photon (upper) and carrier (lower) density distribution at the cavity centre, highlighting their co-localization at the same "hotspot". $N$ is the carrier density and $N_{tr}$ is the transparency carrier density. This strong spatial carrier localization enhances surface recombination, necessitating surface passivation. The concentric "circles" depict a 5 nm $Al_2O_3$ layer used for surface passivation. Scale bar: 100 nm. **d.** Schematics of the cross-section of the central dielectric nanobridge without (upper) and with (lower) the surface passivation. The QWs (red) are first repaired and sealed with InP material (gray), then encapsulated by an $Al_2O_3$ layer (yellow) to mitigate the surface recombination. **e.** SEM images of the central nanobridge without (upper) and with (lower) the surface passivation. Scale bar: 100 nm.

The EDC nanocavity (Fig. 1a-1c) features a sub-diffraction-limit $V_{mod}$ of $0.88(\lambda/(2n))^3$, governed by a central dielectric nanobridge (80 nm width with 70 nm III-V materials). Topology optimization [8] is employed for cavity design to maximize the $Q$-factor while maintaining such a small $V_{mod}$. However, topology optimization is not mandatory; the EDC structure can be adapted to various cavity configurations [36]. Our specific cavity design is chosen to facilitate a fair comparison with a previous state-of-the-art nanolaser (SI, section C). With such a small $V_{mod}$, the carrier distribution, rather than the mode profile, limits the light-matter interaction strength. Therefore, to keep a small $V_I$ with a large optical confinement factor and avoid lateral quantum confinement [37], we do not minimize the bridge width [11] (SI, section E.3). To validate the strategy, we develop a laser model including photon and carrier spatial distributions (SI, section C). Simulations show central localization of both photons and carriers (Fig. 1e), ensuring large spatial overlap and a minimized interaction volume.

The structure is an InP quantum-well (QW) membrane. The fabrication follows the procedure in Ref. [11] with added surface passivation, involving wet-etching for surface preparation, metal-organic-vapor-phase-epitaxy (MOVPE) annealing to reduce surface states and seal the surface, and $Al_2O_3$ deposition to further improve the sealing (Fig. 1d). These shield the QWs from air exposure and smoothens the sidewalls (Fig. 1e). Unlike conventional approaches that chemically treats the surface to remove dangling bonds and then encapsulates with dielectrics [38]-[40] , our method directly seals QWs with InP using MOVPE annealing (Fig. 1d). This technique, similar to buried heterostructure technology [33]-[35], repairs crystal damage caused by ion bombardment during semiconductor dry etching. Additional chemical treatment of the InP sidewalls and final encapsulation with $Al_2O_3$ boost device stability and mitigate long-term degradation. This surface passivation is vital as it effectively suppresses surface recombination, a pervasive issue for nanostructures and particularly important for EDC lasers, where carriers are densely concentrated in space (SI, section B).

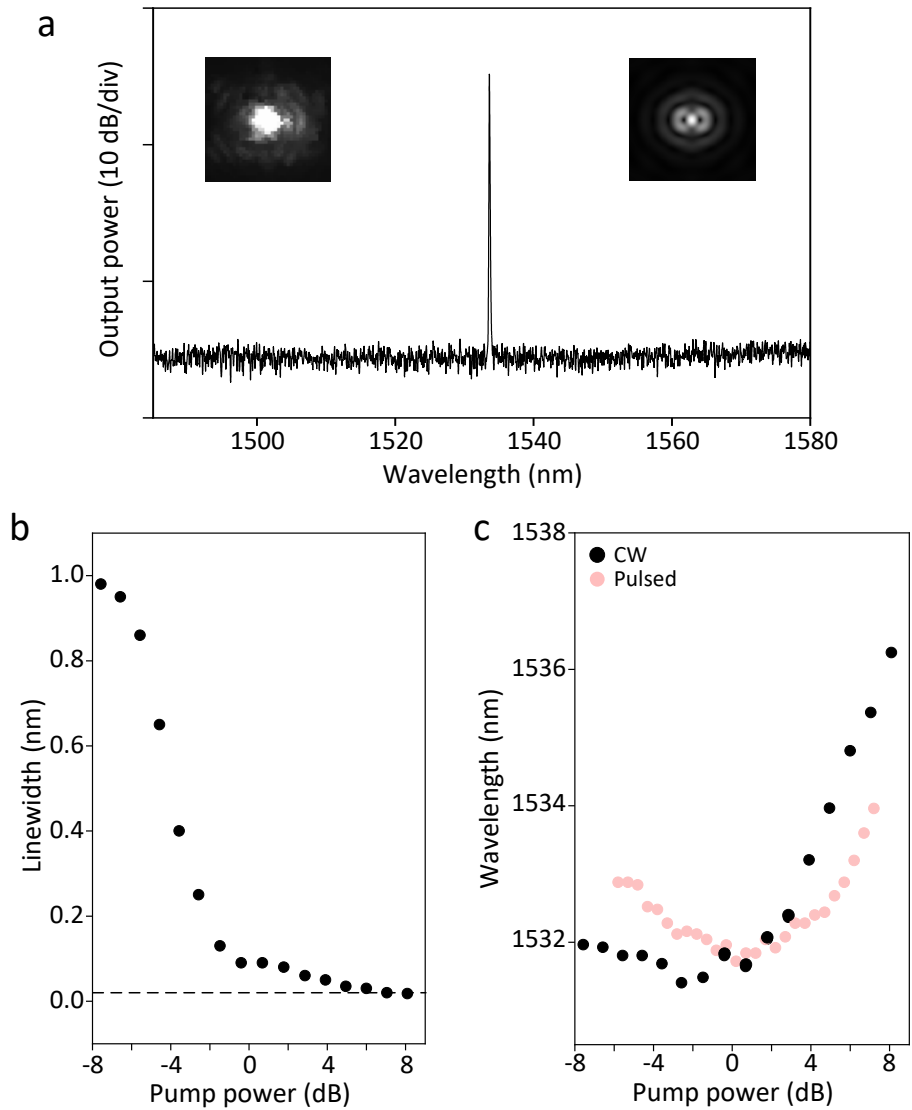

**Fig. 2: Continuous-wave, room-temperature lasing in the EDC laser. a.** Emission spectrum of the EDC laser above threshold, showing single-mode lasing. Insets: measured (left) and simulated (right) emission patterns. **b.**

Linewidth versus pump power, exhibiting a narrow linewidth limited at 0.02 nm (dashed line) by our spectrum analyzer at high pump powers. **c.** Lasing wavelength versus pump power, showing a redshift above threshold. The shift is more pronounced under the CW (black) than pulsed (red) pumping. In (**b**) and (**c**), the pump power is normalized with 0 dB at threshold point.

The device is characterized by optical CW pumping at room-temperature. The EDC laser exhibits single-mode lasing at 1535 nm (Fig. 2a), agreeing with simulations. Lasing is confirmed by the emission pattern (insets in Fig. 2a) and the narrow linewidth, limited at 0.02 nm by the optical spectrum analyzer (Fig. 2b). The input-output measurement displays a typical S-curve but flattens at high pump powers (Fig. 3a), where the lasing wavelength redshifts (Fig. 2c) due to thermal effects. We also performed pulsed pumping at 980 nm, where the redshift is suppressed (red dots in Fig. 2c) (the lasers also work under 980 nm CW pumping).

**Spatial localization of electrons**

We compare the EDC laser to a referenced nanolaser with the same footprint and fabrication process, and a $V_{mod}$ = $2.2(\lambda/2n)^3$, the smallest among PhC point-defect cavities [41], [42]. The total $Q$-factors for the EDC and PhC lasers are estimated at ~6500 and 14000, respectively. Despite expected higher perturbation sensitivity [36], the statistical variations of the EDC lasers are similar to the PhC lasers (shadings in Figs. 3a and 3b), confirming good reproducibility using current nanofabrication technology. An observation is that the EDC laser exhibits a threshold power density of only 5 kW.cm$^{-2}$, significantly lower than the PhC laser despite its lower $Q$-factor. This agrees with simulations, where the modal gain (Fig. 3c), increasing with carrier density, is higher at the device centre below to at threshold, indicating that the lower threshold correlates with a stronger central carrier concentration (Fig. 3d).

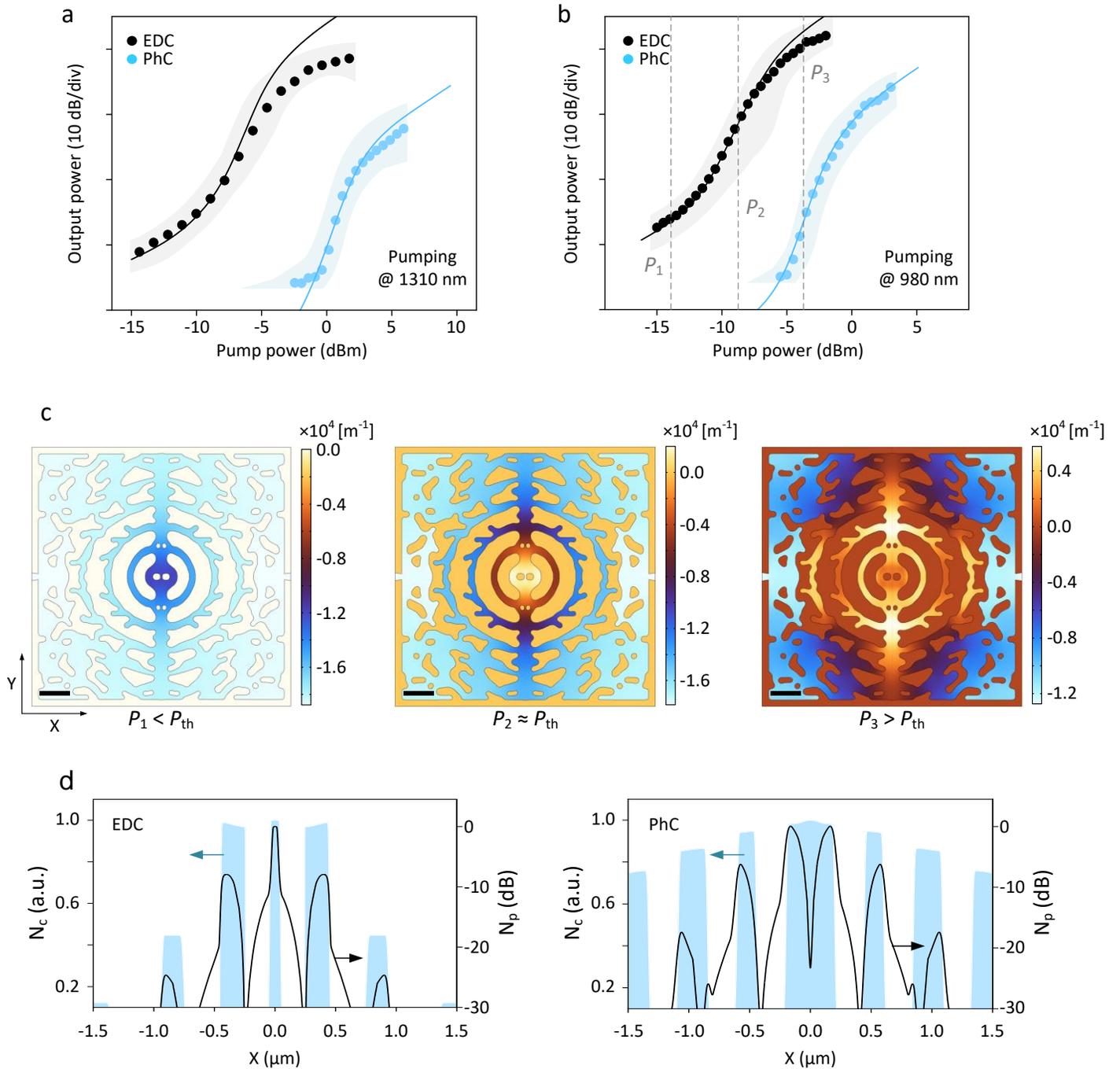

**Fig. 3: Threshold characteristics of the EDC laser. a. b.** Input-output characteristics for the EDC (black) and PhC (blue) laser under 1310 nm CW (**a**) and 980 nm pulsed (**b**) pumping. Experimental results (dots) align closely with theoretical predictions (solid lines), illustrating a lower threshold for the EDC than the PhC laser. The shaded area reflects the statistical variation among 20 nominally identical devices measured across the wafer. **c.** Calculated EDC laser modal gain which increases with carrier density: below (left), near (middle), and above (right) threshold, corresponding to pump powers of $P_1$, $P_2$, and $P_3$, respectively, indicated by the dashed lines in (**b**). Scale bar: 1 μm. **d.** Simulated photon (black curves) and carrier (blue shadings) density distributions at laser threshold under 980 nm pumping for the EDC (left) and PhC (right) laser along the *x*-axis across the device centre. The EDC laser exhibits much stronger field and carrier spatial localization than the PhC laser, with both maxima aligned at the cavity centre. Values are normalized to their respective maxima.

The spatial carrier distribution is shaped by the pumping profile (Fig. 4a). Intriguingly, contrary to what is observed in optical switches where a smaller $V_{mod}$ encourages a faster expansion of the carrier distribution [43], [44], the smaller $V_{mod}$ in the EDC laser enhances carrier localization at the optical "hotspot", fostering a self-alignment of light and matter.

The primary mechanism is that although the pump is a Gaussian beam, the excited field ($|\mathbf{E}_p|^2$) forms an intricate pattern (Fig. 4b) dictated by the electromagnetic boundary conditions and representing a superposition of multiple modes. This leads to not only a more concentrated but also stronger $|\mathbf{E}_p|^2$ in the nanostructured environment than the non-structured case (Fig. 4c). Remarkably, the 980 nm pattern, unlike the much less localized 1310 nm pattern, closely resembles the corresponding lasing mode, with much higher intensity at the EDC cavity centre than the PhC (Fig. 4c). This self-alignment phenomena at shorter wavelengths stems from a smaller spot size and larger light wavevectors, facilitating the excitation of the mode with the smallest $V_{mod}$ through improved spatial and momentum matching. Additionally, our EDC cavity geometry, devoid of competing modes near the pump light frequency, further aids in exciting the lasing mode despite their considerable frequency separation.

Besides the direct benefits from the smaller $V_{mod}$, the EDC structure offers additional ones. Unlike bowtie nanostructures [10], [11], the nanobridge restricts carrier diffusion by reducing the diffusion space from 2D to quasi-1D [43], [44]. Moreover, the EDC cavity features a larger spontaneous emission factor ($\beta$) at the centre than the outer regions (Fig. 1b). Thereby, in contrast to expectations where stronger field localization increases stimulated emission which depletes more carriers, the enhanced $\beta$, stemming from the smaller $V_{mod}$, helps preserve carriers by preventing their loss to non-lasing modes, thus intensifying the spatial carrier imbalance. Consequently, the heightened central carrier concentration boosts the spatially-averaged $\beta$ experienced by the overall carriers, further reducing the laser threshold.

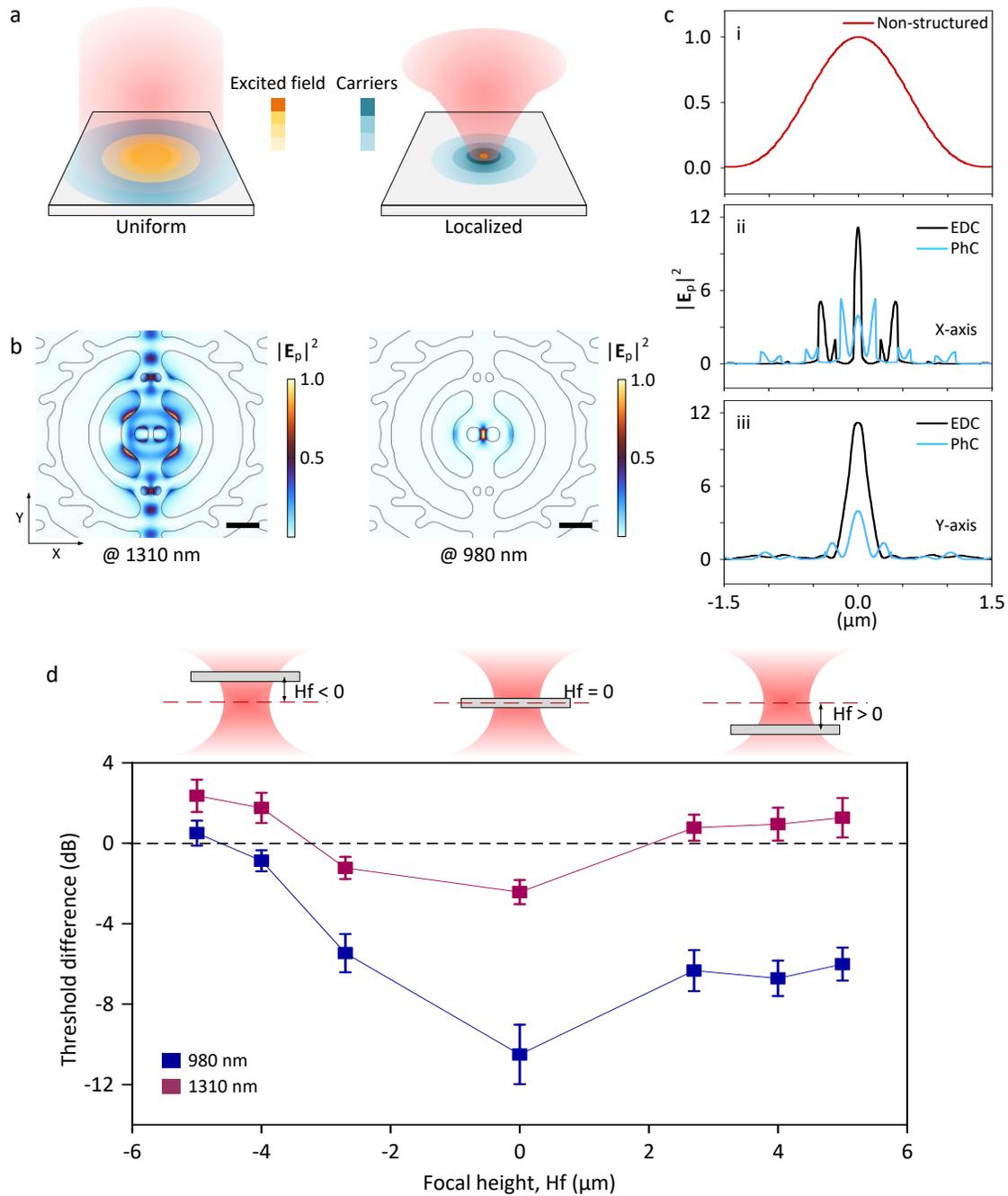

**Fig. 4: Reducing laser threshold via spatial carrier localization. a.** Pumping schemes: Uniform pumping (left) excites multiple cavity modes, resulting in a dispersed field pattern and evenly distributed carriers in space, while localized pumping (right) preferentially excites a mode with a small mode volume, inducing a concentrated carrier distribution. The orange and blue colour bars represent the intensity of the excited field and carriers, respectively,

with darker shades denoting higher intensities. **b.** Simulated pump-excited field ($|E_p|^2$) at 1310 (left) and 980 (right) nm across the EDC laser's central plane, pumped by a focused Gaussian beam as in our experimental setup. Values are normalized to the respective maximum. Scale bar: 0.5 μm, **c.** Simulated pump-excited field energy density ($\varepsilon|E_p|^2$) in a non-structured membrane (i), and in nanostructured cavities along the x-axis (ii) and y-axis (iii) across the centre of the EDC (black) and PhC (blue) lasers. The 980 nm Gaussian beam is used for pumping. Values are normalized to the maximum in (i). $\varepsilon|E_p|^2$ is significantly enhanced at the cavity centre in the EDC compared to both the PhC and non-structured cases, attributable to the electromagnetic boundary conditions enforced by the sharp-edged features of the EDC structure. **d.** Measured threshold difference between the EDC and PhC laser ($\Delta P_{th} = P_{th,EDC} - P_{th,PhC}$) at different pump focal positions. On the left (right), the pump focus is below (above) the membrane, corresponding to a negative (positive) focal height, $H_f$. Blue (purple) data represents $\Delta P_{th}$ for 980 (1310) nm pump, where $|E_p|^2$ is more (less) localized in the EDC than in the PhC laser. Error bars indicate measurement uncertainty across multiple devices.

To assess the carrier localization effect, we measure the laser thresholds by varying the pump focal plane. While the pump power in Fig. 3 represents pre-injection levels, the power absorbed differs between structures due to different excitation patterns. For clarity, we normalize the input power to ensure equal energy absorption for both lasers (SI, section E.2). At the focal plane, the lasers reach their minimum thresholds, which increase as the focal plane shifts since their excitation patterns disperse (SI, section E.2). The EDC laser exhibits a lower threshold, with a reduction of up to 12 (3) dB at the focal plane under the 980 (1310) nm pumping compared to the PhC laser (Fig. 4d). Notably, the EDC laser maintains a lower threshold at 1310 nm than the PhC laser despite its more diffused excitation pattern (Fig. S6 in the SI), highlighting its inherent superiority in carrier localization.

**Discussions**

Given the importance of carrier localization, we introduce the carrier volume $V_{car} = \int N_c(\mathbf{r})d\mathbf{r} / \max\{N_c(\mathbf{r})\}$ to quantify the spatial extent of "matter". The EDC laser achieves a $V_{car}$ of $0.28(\lambda/n)^3$ and $0.4(\lambda/n)^3$ at threshold under the focused 980 nm and 1310 nm pumping, compared to $2.1(\lambda/n)^3$ and $1.45(\lambda/n)^3$ for the PhC laser. This improvement far surpasses what is seen in $V_{mod}$. The effective diameter of $V_{car}$ (if treated as a disc) for the PhC laser is ~3.8 (3.2) µm under 980 (1310) nm pumping, whereas it reduces to ~1.4 (1.7) µm in the EDC laser, which is even smaller than that of the pump beam, measured to be ~2 (2.5) µm.

To understand how the interaction volume $V_I$ depends on $V_{mod}$ and $V_{car}$, we assume that $N_p(\mathbf{r})$ and $N_c(\mathbf{r})$ have in-plane Gaussian distributions. Then, $V_I$ is found to be a combination of $V_{mod}$ and $V_{car}$ (SI, section D.3):

$$V_I = V_{mod} + V_{car}/\Gamma. \tag{3}$$

Normally, $V_{car}$ is much larger than $V_{mod}$ in nanocavities; reducing $V_{car}$ is thus more important. From Eq. (1), the $V_I$ for the EDC laser is $4.2(\lambda/n)^3$ and $5.5(\lambda/n)^3$ under 980 and 1310 nm pumping, compared to $31(\lambda/n)^3$ and $22(\lambda/n)^3$ for the PhC laser. These values align well with the estimates from Eq. (3). The results confirm that the EDC structure enhances carrier concentration in regions with the strongest mode field, effectively reducing $V_I$ which in turn lowers the laser threshold.

We have demonstrated a room-temperature continuous-wave nanolaser exploiting extreme spatial localization of both light and excited electrons at the same "hotspot". This self-alignment leads to enhanced light-matter interactions and a significant reduction in the laser threshold. We introduce the light-matter interaction volume as the key metric to quantify this effect. Compared to other types of lasers, the EDC laser effectively manages the traditional trade-off between carrier volume and available electron states, similar to how it addresses the trade-off between mode volume and cavity $Q$-factor. The EDC thus substantially lowers the interaction volume while maintaining high gain and cavity $Q$-factor. This advancement is expected to elevate light-matter interactions to unprecedented high levels. The principle extends beyond light sources, impacting other optoelectronic devices [45]-[49], as well as sensing and imaging [50], [51].

## Methods

### Fabrication process

The device is fabricated within a 250 nm thick InP membrane embedded with three InGaAsP quantum-well (QW) layers. Firstly, the InP wafer is flip-bonded to a Si/SiO$_2$ substrate. The laser membrane is formed after removing the InP substrate and an InGaAs etch-stop layer.

A SiN$_x$ layer is deposited onto the wafer, followed by the spin-coating of a photoresist. The structures are then patterned using electron-beam lithography. This pattern is transferred to the SiN$_x$ hard mask and subsequently to the InP layer through a two-step semiconductor etching. Following the etching, the sample undergoes a passivation process, where it is first treated in a solution of ammonium hydroxide to remove the oxide layer and then annealed inside the metal-organic-vapour-phase-epitaxy reactor to repair and seal the QWs. After the annealing, the structures are membranized and encapsulated with an Al$_2$O$_3$ layer via atomic layer deposition. Further details are provided in section B in the Supplementary Information.

### Experimental setup

The samples are vertically pumped by a Gaussian light beam through an objective lens with a numerical aperture of 0.65 using a micro-photoluminescence setup. This setup contains three laser diodes (Thorlabs CLD1015) as the pump source: a continuous-wave laser at 1310 nm or 980 nm, or a pulsed 980 nm laser. The laser is cascaded with an attenuator to adjust the pump power. Precise control over the pump position and area is maintained and monitored by an infrared camera (Xeva). The emission from the sample is collected vertically using the same objective lens. For the output measurements, after being isolated from the reflected pump beam by a long-pass filter, the collected signal is analyzed using an optical spectrum analyzer (YOKOGAWA AQ6370D) with a spectral resolution of 0.02 nm.

As for the pulsed pumping, we chose a pulse width of 0.2 µs with a duty cycle of 10 %. This pulse width ensures that the transient regime, during which the laser reaches its steady state, is short compared to the pulse width so that the lasing state is maintained for an extended period. The low duty cycle effectively reduces thermal effects. To adjust the focal plane of the pump light, we tune the voltage of a piezo actuator (Thorlabs MDT693A) correlated with the *z*-direction of an XYZ stage (NanoMax-TS) holding the sample. This stage provides a movement distance of 20 µm per 75 V applied.

**Numerical simulation and theory**

The optical properties of the devices are calculated using both finite-difference time-domain (FDTD) simulations based on Ansys Lumerical FDTD and finite element method (FEM) calculations based on COMSOL Multiphysics. The two methods show good agreement. Additionally, a two-dimensional QW laser model incorporating spatial details of the optical mode field, pump-excited field pattern, spontaneous emission factor, and carrier transport is developed based on FEM using COMSOL Multiphysics. Further details on the model and underlying theory are provided in sections C and D in the Supplementary Information.

**Data availability**

All data in this study are available within the paper and its Supplementary Information. Further source data will be made available on reasonable request.

**Acknowledgements**

The authors acknowledge R. E. Christiansen, F. Wang, and O. Sigmund for their design support and G. Dong, B. Munkhbat, and N. Gregersen for their experimental support. This work was supported by the Danish National Research Foundation (Grant no. DNRF147 NanoPhoton), the European Research Council (ERC) under the European Union Horizon 2020 Research and Innovation Programme (Grant no. 834410 FANO), and the Villum Fonden via the Young Investigator Program (Grant no. 42026 EXTREME).

**Author contributions**

M.X. developed the semiconductor etching process and led the fabrication, characterization, and device measurement. Y.B., supervised by E.S. and K.Y., developed the surface passivation process and performed the quantum-well wafer epitaxy. S.K.B., A.H.D., and Y.Y. assisted with the fabrication, characterization, and measurements. Y.Y. and J.M. conceived the concept, and Y.Y. developed the theory and numerical model. Y.Y. and M.X. performed the simulations and analyzed the results. J.M. and Y.Y. supervised the project. Y.Y. and M.X. prepared the manuscript with inputs from J.M. All authors discussed the results and contributed to the final manuscript.

**Competing interests**

The authors declare no competing interests.

# Supplementary Information: A nanolaser with extreme dielectric confinement

Meng Xiong[1,2]†, Yi Yu[1,2]*†, Yury Berdnikov[1], Simon Klinck Borregaard[1,2], Adrian Holm Dubré[1], Elizaveta Semenova[1,2], Kresten Yvind[1,2], and Jesper Mørk[1,2]*

[1]Department of Electrical and Photonics Engineering, Technical University of Denmark, Ørsteds Plads 345A, 2800 Lyngby, Denmark.

[2]NanoPhoton - Center for Nanophotonics, Ørsteds Plads 345A, 2800 Lyngby, Denmark.

†These authors contribute equally to this work.

*Corresponding author: yiyu@dtu.dk; jesm@dtu.dk

## A. Nanolasers with sub-diffraction-limit mode volumes

Table S1 lists representative nanolasers with sub-diffraction-limit mode volumes. The extreme dielectric confinement (EDC) laser stands out with its high quality-factor ($Q$-factor), large gain and optical confinement factor, and more importantly, an ultrasmall interaction volume, enabling continuous-wave (CW) lasing at room-temperature.

Table S1. Nanolasers with sub-diffraction-limit mode volumes*

| Laser type | Mode volume | Lasing wavelength | Threshold pump power density | Operating conditions |
|---|---|---|---|---|
| Plasmonic [1] | $0.000015\lambda^3$ | 373 nm | >50 MW.cm$^{-2}$ | Pulsed (cryostat) |
| Plasmonic [2] | $0.0004(\lambda/(2n))^3$ | 870 nm | 7 GW.cm$^{-2}$ | Pulsed (RT) |
| Plasmonic [3] | $0.56(\lambda/(2n))^3$ | 1308 nm | 120 kW.cm$^{-2}$ | Pulsed (cryostat) |
| Metallic [4] | $0.5(\lambda/(2n))^3$ | 1354 nm | 71 kW.cm$^{-2}$ | Pulsed (RT) |
| Metallic [5] | $0.001\lambda^3$ | 1550 nm | 20 kW.cm$^{-2}$ | Pulsed (RT) |
| Dielectric void [6] | $0.0005\lambda^3$ | 1580 nm | 26 kW.cm$^{-2}$ | Pulsed (RT) |
| Dielectric void [7] | $0.003\lambda^3$ | 1550 nm | >2 kW.cm$^{-2}$ | Pulsed (RT, in water) |
| EDC (this work) | $0.003\lambda^3$ | 1535 nm | 5 kW.cm$^{-2}$ | CW (RT) |

*The table is compiled from literature reporting mode volumes below the diffraction limit (($\lambda/(2n))^3$). Notably, a metallic nanolaser has demonstrated continuous-wave operation at room-temperature [8], but its mode volume of $5.88(\lambda/(2n))^3$ exceeds the diffraction limit. The pump power refers to the injected power before impinging on the device and is the peak-level power for pulsed pumping. For plasmonic/metallic lasers, the effective index ($n$) is often unspecified (e.g., Refs [2]-[4]) and depends on the mode profile. RT: room-temperature.

## B. Device fabrication

Figure S1 illustrates the fabrication process. The device, with a ~7 μm×7 μm footprint, is fabricated within a 250 nm thick InP membrane embedded with three layers of InGaAsP/AlGaAs quantum wells (QWs). The InP wafer, with a 100 nm InGaAs etch-stop layer, is grown on an InP substrate via metal-organic-vapor-phase-epitaxy (MOVPE). Firstly, the wafer is flip-bonded to a Si/SiO$_2$ substrate. The Si/SiO$_2$ substrate is initially prepared through oxidation and annealing of a Si wafer in an anneal-oxide furnace at 1100 °C, yielding a 1.1 μm thick SiO$_2$ layer, which serves as a sacrificial layer for the membranization process. Before bonding, the surfaces of the InP wafer and the Si/SiO$_2$ substrate are activated with O$_2$ plasma for 30 seconds. The InP wafer is then directly bonded to the Si/SiO$_2$ substrate using a Süss SB6 wafer bonder at 300°C under a force of 2 kN. The InP substrate and InGaAs etch-stop layer are then removed to form the QW membrane.

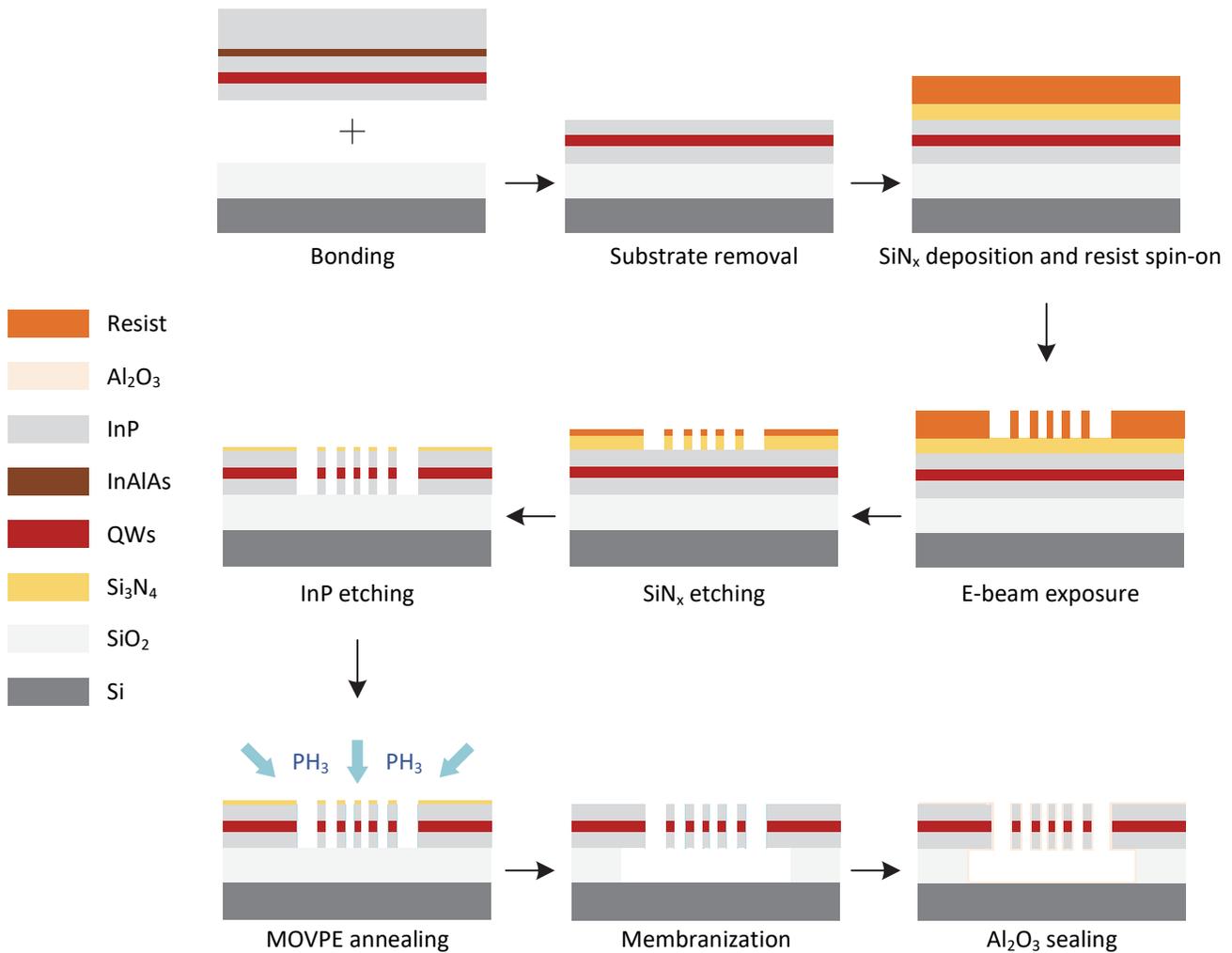

Fig. S1. Schematic of the device fabrication process.

A 100 nm thick SiN$_x$ layer is deposited onto the wafer, followed by a 180 nm thick chemically semi-amplified resist (CSAR 6200.09) via spin-coating. The SiN$_x$ and CSAR layers serve as a hard mask and photoresist, respectively, with thicknesses optimized to allow etchant access to narrow cavity openings while protecting the underlying materials.

The laser structures are patterned using a JEOL-9500FSZ electron-beam writer with a 0.2 nA current and 1 nm shot pitch for high resolution. This pattern is first transferred to the SiN$_x$ hard mask, then to the InP layer through a two-step inductively coupled plasma (ICP) etching. During ICP etching, the sample is mounted on a Si carrier wafer. This carrier wafer can form a passivation layer when interacting with the plasma, helping accumulate material on the sidewalls and effectively reducing lateral etching [9]. For InP etching, HBr gas is employed at a controlled flow rate of 5 sccm to maintain a low pressure of 0.5 mTorr. This low pressure is crucial for achieving straight sidewalls.

Following the etching, the sample undergoes passivation. The sample is first treated with ammonium hydroxide to remove the oxide layer formed on the sidewalls of the cavities during the etching and contact with the atmosphere. Then, the sample is annealed for 10 min under phosphine flux at 600°C inside the MOVPE reactor. At this temperature, annealing under phosphine initiates the replacement of arsenic atoms with phosphorus near the sidewall surface, thus isolating the QWs with a phosphorus-rich shell. After the annealing, the structures are membranized using a buffered hydrofluoric acid and then sealed with a 5 nm-thick layer of Al$_2$O$_3$ via thermal atomic layer deposition inside the Picosun R200 system. Compared to conventional methods that rely solely on chemical treatments followed by dielectric encapsulation [17]-[19], our surface passivation process incorporates an additional step of crystal quality repair and sealing of the InGaAsP QWs with "InP" through MOVPE annealing. This approach could be more effective at suppressing surface recombination (see section C.2.3).

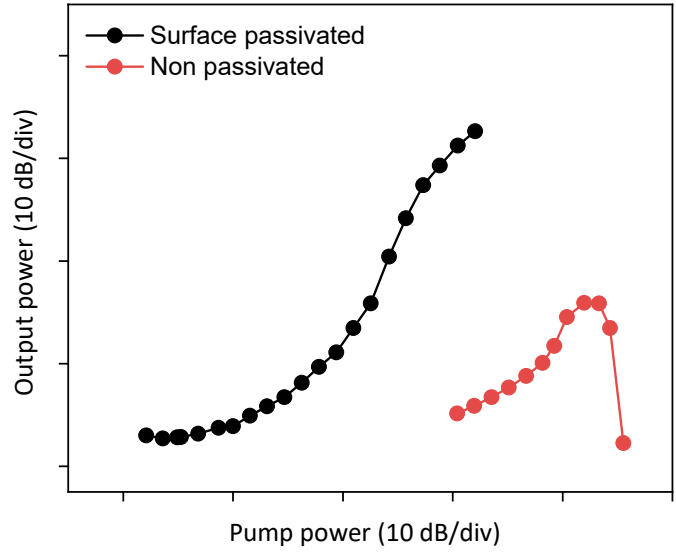

Fig. S2. Examples of measured input-output characteristics of EDC lasers with (black) and without (red) surface passivation. The abrupt decrease in output power at alleviated pump power levels in the non-passivated EDC laser is due to the damage of the central dielectric nanobridge.

Surface recombination is a common issue for nanostructures with high surface-to-volume ratios, leading to enhanced carrier losses and exacerbating heating problems. Prior experiments on similar EDC lasers without surface passivation required pulsed operation at room-temperature or CW operation at cryogenic temperatures. In contrast, the photonic-crystal (PhC) point-defect (H0-type) nanocavity lasers (and other types of PhC lasers) can exhibit CW lasing at room-temperature without surface passivation, though their performance improves with it. Room-temperature CW pumping of unpassivated EDC cavities either failed to produce lasing or caused permanent damage (Fig. S2), with subsequent inspections revealing burning of the dielectric nanobridge, in agreement with the strong carrier localization there.

## C. Numerical model

### C.1. Estimation of the spontaneous emission factor

For a quantum emitter at position $\mathbf{r}_0$ interacting with a cavity mode $i$, the emission decay rate is determined using Fermi's golden rule [10]

$$\Gamma_{e,i}(\mathbf{r}_0) = \frac{2\pi}{\hbar^2}\left|\mathbf{E}_i(\mathbf{r}_0)\cdot\vec{d}\sqrt{n_p+1}\right|^2 \int_{-\infty}^{\infty} d\omega D_c(\omega)L(\omega), \tag{S.1}$$

where $\vec{d}$ is the dipole moment of the quantum emitter, $n_p$ is the total number of photons, $\mathbf{E}_i(\mathbf{r}_0) = E_i(\mathbf{r}_0)\vec{e}$ ($\vec{e}$ is the unit vector) is the vacuum electrical field, with an amplitude $E_i(\mathbf{r}_0) = \sqrt{\hbar\omega_c/(2\varepsilon(\mathbf{r}_0)V(\mathbf{r}_0))}$, and

$$V(\mathbf{r}_0) = \int \varepsilon(\mathbf{r})|E_i(\mathbf{r})|^2 d\mathbf{r} / \left(\varepsilon(\mathbf{r}_0)|E_i(\mathbf{r}_0)|^2\right).$$

Here, $\varepsilon(\mathbf{r}) = \varepsilon_0\varepsilon_r(\mathbf{r})$ with $\varepsilon_r(\mathbf{r})$ being the relative permittivity. For spontaneous emission where $n_p = 0$, we can derive from Eq. (S.1) the following expression:

$$\Gamma_{e,i}(\mathbf{r}_0) = \frac{2\omega_c}{2\hbar\varepsilon(\mathbf{r}_m)V(\mathbf{r}_m)}\left|\vec{e}\cdot\vec{d}\right|^2 \frac{(\gamma_c+\gamma_e)/2}{(\omega_c-\omega_e)^2+((\gamma_c+\gamma_e)/2)^2}\frac{|E_i(\mathbf{r}_0)|^2}{|E_i(\mathbf{r}_m)|^2} = 2g_c^2 \cos(\phi)\cos(\varphi)\rho(\omega_c). \tag{S.2}$$

In this formula, $\omega_c$ and $Q$ are the resonant frequency and $Q$-factor of the cavity mode $i$, $\gamma_c = \omega_c/Q$ is the inverse of the cavity photon lifetime, $\gamma_e$ represents the bandwidth of the emitter and $\omega_c - \omega_e$ is the detuning of $\omega_c$ with respect to the centre frequency of the emitter $\omega_e$. The quantum emitter models the bandwidth and peak of our QW photoluminescence spectrum. The parameter $g_c = d\sqrt{\hbar\omega_c/(2\varepsilon(\mathbf{r}_m)V_{\mathrm{mod}})}/\hbar$ is the conventional coupling strength between the cavity mode and the quantum emitter, in which $\mathbf{r}_m$ is the position where $\varepsilon(\mathbf{r})|E_i(\mathbf{r})|^2$ is maximized, and $V_{\mathrm{mod}} = \int \varepsilon(\mathbf{r})|E_i(\mathbf{r})|^2 d\mathbf{r} / \left(\varepsilon(\mathbf{r}_m)|E_i(\mathbf{r}_m)|^2\right)$ is the optical mode volume. The factor $\cos(\phi) = |\vec{e}\cdot\vec{d}|^2/d^2$ accounts for imperfect alignment of quantum emitter polarization relative to the polarization of the local electrical field, and $\cos(\varphi) = |E_i(\mathbf{r}_0)|^2/|E_i(\mathbf{r}_m)|^2$ accounts for spatial offset between the emitter and the mode field intensity maximum, and $\rho_e(\omega_c) = (\gamma_c+\gamma_e)/\left(2\left((\omega_c-\omega_e)^2+((\gamma_c+\gamma_e)/2)^2\right)\right)$ accounts for detuning between the cavity spectrum and the active material.

Eq. (S.2) describes the decay rate of a single emitter. To derive the macroscopic decay rate, one needs to consider the QW local electronic density of states and the occupation probabilities of electrons and holes, determined by excited carrier density and, hence, the pump strength [11]. Here, we use Eq. (S.2) as a reasonable approximation for moderate carrier density. While the specific form of the decay rate may vary, the ratios between different cavity modes typically remain more constant. Thus, we approximate the spontaneous emission factor (for mode $i$ in 2D space) by

$$\beta_i(x,y) = \Gamma_{e,i}(x,y) / \sum_{n=1}^{\infty} \Gamma_{e,n}(x,y). \quad (S.3)$$

It is important to note that in this study, $\beta_i$ refers to the "optical" type, which excludes nonradiative recombination. In our implementations, $\beta_i$ is obtained numerically from a finite series of quasi-normal modes calculated within a 150 nm wavelength range centred on the targeted lasing mode $i$.

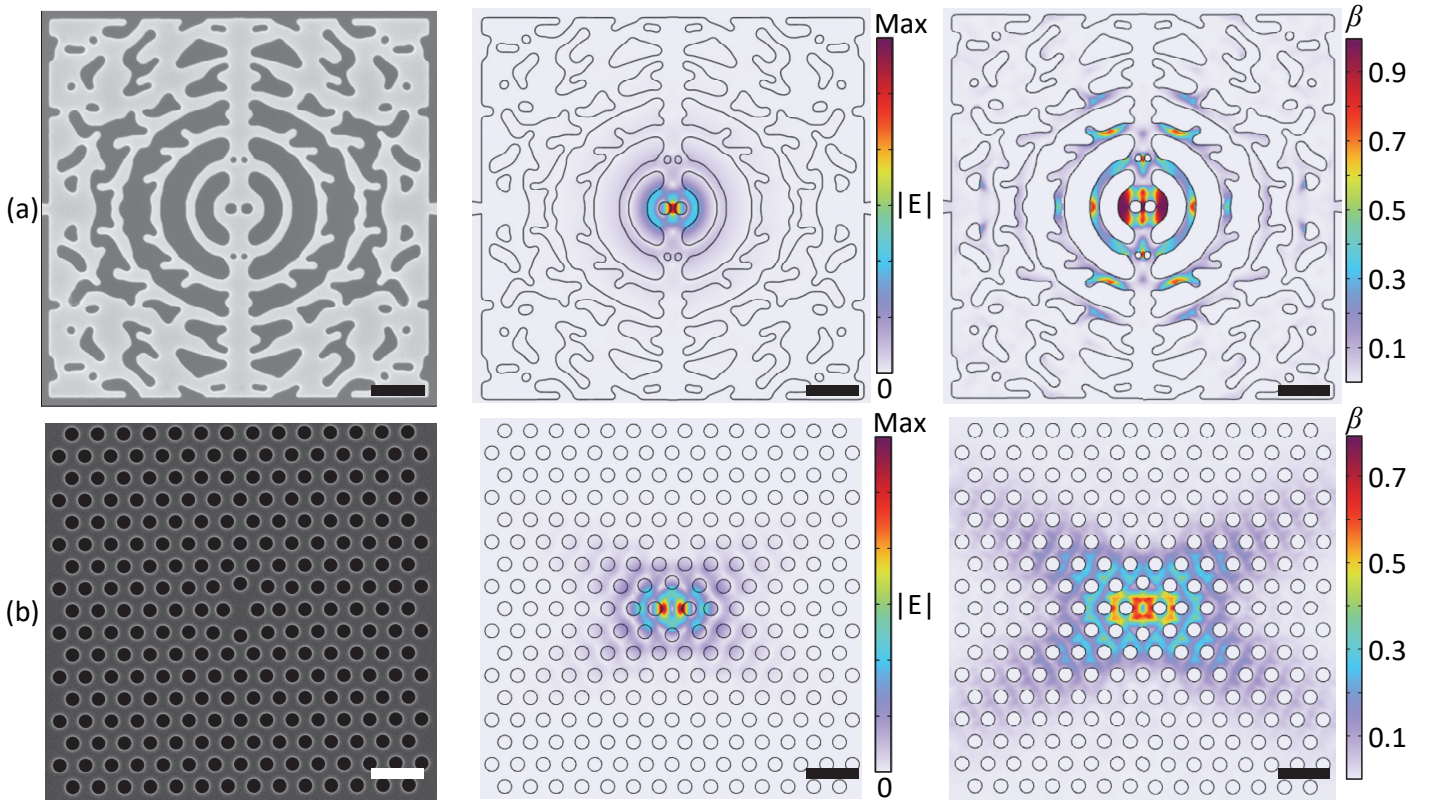

Fig. S3. (a, b) SEM images (left) of the EDC (a) and PhC H0 (b) cavities, along with the spatial profiles of their corresponding modes (middle) and spontaneous emission factors based on Eq. (S.3) (right). Scale bar: 1 μm.

The PhC H0 nanocavity is selected as a reference for its superior characteristics [12], including its small $V_{mod}$ and high $Q$-factor for such a compact footprint, which are challenging to achieve with other designs. The PhC has the same membrane thickness as the EDC cavity (a thinner membrane leads to a smaller $V_{mod}$). Besides, both cavities lack modes near our pump light frequencies, ensuring undisturbed pumping efficiencies. Additionally, they share the same footprint, similar optical confinement factors, dielectric-to-air in-plane area ratios (impacting pumping efficiency), sidewall dielectric-air interface areas (influencing surface recombination), and comparable $Q$-factors.

Fig. S3 shows that the EDC cavity not only has a smaller $V_{mod}$ but also a larger $\beta_i$ at the centre compared to the PhC. This enhanced $\beta_i$ is primarily due to only two major modes (modes having high intensity) at the EDC centre, whereas the H0 nanocavity features three modes. A smaller mode volume implies a larger free spectral range and fewer modes within a specified distance around the point where the mode volume is evaluated.

### C.2. Two-dimensional quantum-well laser model

We present our two-dimensional QW laser model, which, unlike previous efforts [13], incorporates spatial variations in the spontaneous emission factor, pump-induced excitation pattern along with carrier diffusion.

### C.2.1. Carrier rate equation

The carrier density $N_{ca}(\mathbf{r},t)$ in the QW can be separated as

$$N_{ca}(\mathbf{r},t) = N(x,y,t) N_z(z). \tag{S.4}$$

Here, $N(x,y,t)$ ($N_z(z)$) is the carrier density across the QW plane (along the QW growth direction). By assuming $N_z(z)$ uniform and normalized such that $N_z(z)=1$, we get $N_{ca}(\mathbf{r},t) = N(x,y,t)$. The total carrier number, $n_c(t)$, is then given by

$$n_c(t) = \int N_{ca}(\mathbf{r},t) d\mathbf{r} = \int_z N_z(z) dz \iint_{xy} N(x,y,t) dxdy = h_{QW} \iint_{xy} N(x,y,t) dxdy. \tag{S.5}$$

Here, $h_{QW}$ is the total QW thickness. Our samples use three QWs, each about 7 nm thick. To avoid accounting for the vertical variations in field and carrier density, which is a minor effect, we consider a single effective active layer embedded at the device's central plane with an effective thickness of 20 nm.

The variable separation in Eq. (S.4) simplifies the problem to being two-dimensional, enabling us to focus solely on $N(x, y, t)$, which can be described by the following rate equation

$$\frac{d}{dt} N(x,y,t) = \frac{1}{q} \nabla \cdot J(x,y,t) - R(x,y,t) + G_p(x,y,t), \tag{S.6}$$

where

$$J(x,y,t) = q\mu N(x,y,t) \mathbf{E}_b(x,y,t) + qD\nabla N(x,y,t). \tag{S.7}$$

Here, $\mu$ and $D$ are the carrier drift and diffusion coefficients, $q$ is the electron charge, and $\mathbf{E}_b(x,y,t)$ is the background electrical field. Focusing on the laser steady-state, we adopt the ambipolar approximation [14], [15]. Neglecting the drift effects, which are minor due to field screening by excited carriers, one gets

$$\nabla \cdot J(x,y,t) = qD\nabla^2 N(x,y,t).$$

In Eq. (S.6), $G_p(x,y,t)$ is the carrier generation rate, which scales with the pump power, and has a pattern determined by the pump source. $R(x,y,t)$ is the carrier recombination rate, which consists of stimulated emission, spontaneous emission, and nonradiative recombination, specifically

$$R(x,y,t) = R_{st}(N(x,y,t)) + R_{sp}(N(x,y,t)) + R_{nr}(N(x,y,t)). \tag{S.8}$$

The stimulated emission rate can be derived as

$$R_{st}(N(x,y,t)) = G_N(N(x,y,t)) n_p(t) |\mathbf{E}_{in}(x,y)|^2, \tag{S.9}$$

where we employ a QW gain model [11]:

$$G_N(N(x,y,t)) = \frac{\Gamma_z}{h_{QW}} v_g g_0 \ln\left(\frac{N(x,y,t) + N_s}{N_{tr} + N_s}\right).$$

Here, $n_p(t)$ is the photon number, $|\mathbf{E}_{in}(x,y)|^2$ is the in-plane field energy density, $\Gamma_z$ is the vertical ($z$ direction) optical confinement factor. Their details are given in the next section. Additionally, $\upsilon_g$ is the material group velocity, $g_0$ is the material gain coefficient, $N_{tr}$ is the transparency carrier density, and $N_s$ is the so-called "linear parameter" employed to fit experimental data. The total spontaneous emission rate can be derived as

$$R_{sp}(N(x,y,t)) = \frac{R_{sp,i}(N(x,y,t))}{\beta_i(x,y)} = \frac{1}{\beta_i(x,y)} R_N(N(x,y,t)) |\mathbf{E}_{in}(x,y)|^2, \quad \text{(S.10)}$$

where an empirical formula [16] is used

$$R_N(N(x,y,t)) = \frac{1}{2} \frac{\Gamma_z}{h_{\text{QW}}} \upsilon_g g_0 \ln\left(1 + \left(\frac{N(x,y,t) + N_s}{N_{tr} + N_s}\right)^2\right).$$

$R_{sp,i}(N(x,y,t))$ is the spontaneous emission rate into the lasing mode (mode $i$), and $\beta_i(x,y)$ is the spontaneous emission factor presented in section C.1. The nonradiative recombination can be expressed by

$$R_{nr}(N(x,y,t)) = \gamma_L N(x,y,t) + \gamma_A N(x,y,t)^3 + R_{sr}(N(x,y,t)),$$

in which $\gamma_L N(x,y,t)$ represents the linear recombination and $\gamma_A N(x,y,t)^3$ reflects the Auger process. $R_{sr}(N(x,y,t))$ denotes surface recombination, occurring exclusively at dielectric-air interfaces. This surface recombination necessitates the Neumann boundary condition: $D\nabla N(x,y,t) = -SN(x,y,t)$, where $S$ is the surface recombination velocity.

### C.2.2. Photon rate equation

Similarly, the photon density $N_{ph}$ may be decomposed as

$$N_{ph}(\mathbf{r},t) = n_p(t)\varepsilon(\mathbf{r})|\mathbf{E}(\mathbf{r})|^2 \approx n_p(t)\varepsilon(x,y)|\mathbf{E}_z(z)|^2|\mathbf{E}_{xy}(x,y)|^2 = n_p(t)|\mathbf{E}_z(z)|^2|\mathbf{E}_{in}(x,y)|^2.$$

We use $\varepsilon(\mathbf{r}) \approx \varepsilon(x,y)$ given that the mode is typically well confined within the membrane and we neglect the difference between the refractive indices of the QW region and the cladding layers. A simplified notation, $|E_{in}(x,y)|^2 = \varepsilon(x,y)|E_{xy}(x,y)|^2$ is adopted. The total photon number is

$$\int N_{ph}(\mathbf{r},t)d\mathbf{r} = n_p(t)\int N_P(\mathbf{r})d\mathbf{r} = n_p(t)\int_z |E_z(z)|^2 dz \iint_{xy} |E_{in}(x,y)|^2 dxdy = n_p(t). \quad (S.11)$$

The field has been normalized such that $\int_z |E_z(z)|^2 dz = 1$ and $\iint_{xy} |E_{in}(x,y)|^2 dxdy = 1$. The in-plane optical confinement factor is defined as $\Gamma_{xy} = \int_{xy \in QW} |E_{in}(x,y)|^2 dxdy / \int_{xy} |E_{in}(x,y)|^2 dxdy$, which is found to be 0.93 (0.96) for our EDC (PhC H0) nanocavity. Noting that this definition of the optical confinement factor (where the integration region extends over the entire membrane containing QWs) is general and encompasses cases where the active region has defined lateral boundaries, as seen in conventional lasers with buried heterostructures [11]. By employing the relations ($z_w$ denotes the active layer in $z$, i.e., the centre plane of the membrane)

$$\int_{z \in QW} |E_z(z)|^2 dz = \int_{z \in QW} |E_z(z)|^2 dz / \int_z |E_z(z)|^2 dz \approx h_{QW} |E_z(z_w)|^2 = \Gamma_z,$$

one gets $|E_z(z_w)|^2 = \Gamma_z / h_{QW}$, and the dynamics of photon number can be derived as

$$\frac{d}{dt}n_p(t) = -\gamma_c n_p(t) + \frac{h_{QW}}{\Gamma_z} n_p(t) \int G_N(N(\mathbf{r},t)) \varepsilon(\mathbf{r})|E(\mathbf{r})|^2 d\mathbf{r} + \frac{h_{QW}}{\Gamma_z} \int R_N(N(\mathbf{r},t)) \varepsilon(\mathbf{r})|E(\mathbf{r})|^2 d\mathbf{r}$$
$$\approx -\gamma_c n_p(t) + h_{QW} n_p(t) \iint_{xy \in QW} G_N(N(x,y,t))|E_{in}(x,y)|^2 dxdy + h_{QW} \iint_{xy \in QW} R_N(N(x,y,t))|E_{in}(x,y)|^2 dxdy$$

(S.12)

Since $\varepsilon(\mathbf{r})|E(\mathbf{r})|^2$ equals to $N_p$, and $G_N$ and $R_N$ are proportional to $N$ (and $N_c = N/n_c$ as shown in the following section), the spatial integral in Eq. (S.12) is proportional to the volume integral of $\int N_p(\mathbf{r}) N_c(\mathbf{r}) d\mathbf{r}$, which is the inverse of the interaction volume to be discussed in Section D.

## C.2.3. Implementation and parameters

Simulations are conducted as follows: First, the pumping profile $G_P(x,y)$ is calculated through FDTD (see section E.1). Subsequently, the lasing mode and $Q$-factor are computed via FDTD and FEM (eigenmode solver), followed by the calculation of $\beta_i$ using FEM. These values are treated as constants in Eqs. (S.6) and (S.12), which are solved numerically using FEM. The parameters, comparable to those of conventional semiconductor QW lasers [11], are listed in Table S2. Notably, a surface recombination velocity of $S = 2500$ cm/s is extracted from theoretical fitting, much lower than the state-of-the-art values [17], [18]. Ref. [19] achieved an ultra-low surface recombination velocity of only 260 cm/s, but this was with a much thicker dielectric capping layer of over 50 nm. The same study showed that this velocity rises to approximately 7000 cm/s with a thinner 5 nm (as in our case) dielectric capping layer.

Table S2. Parameter values used in simulations

| Parameter | Symbol | Value |
|---|---|---|
| Refractive index | $n$ | 3.2(InP), 1.75(Al$_2$O$_3$) |
| Diffusion coefficient | $D$ | $5.75 \times 10^{-4}$ m$^2$.s$^{-1}$ |
| Surface recombination velocity | $S$ | 25 m.s$^{-1}$ |
| Group velocity | $v_g$ | $9.4 \times 10^7$ m.s$^{-1}$ |
| Total QW thickness (3 QWs) | $h_{QW}$ | 20 nm |
| Gain coefficient | $g_0$ | 1800 cm$^{-1}$ |
| Transparency carrier density | $N_{tr}$ | $1.2 \times 10^{24}$ m$^{-3}$ |
| Carrier linear parameter | $N_s$ | $0.6 \times 10^{24}$ m$^{-3}$ |
| Linear recombination coefficient | $\gamma_L$ | $3.3 \times 10^7$ s$^{-1}$ |
| Auger recombination coefficient | $\gamma_A$ | $5 \times 10^{-41}$ m$^6$.s$^{-1}$ |
| Intrinsic $Q$-factor | $Q_i$ | 11000(EDC), 93000(PhC) |
| Absorption $Q$-factor | $Q_a$ | 16000 |
| Total $Q$-factor | $Q$ | $1/(1/Q_i+1/Q_a)$ |
| Vertical optical confinement factor (3 QWs) | $\Gamma_z$ | 0.096 |

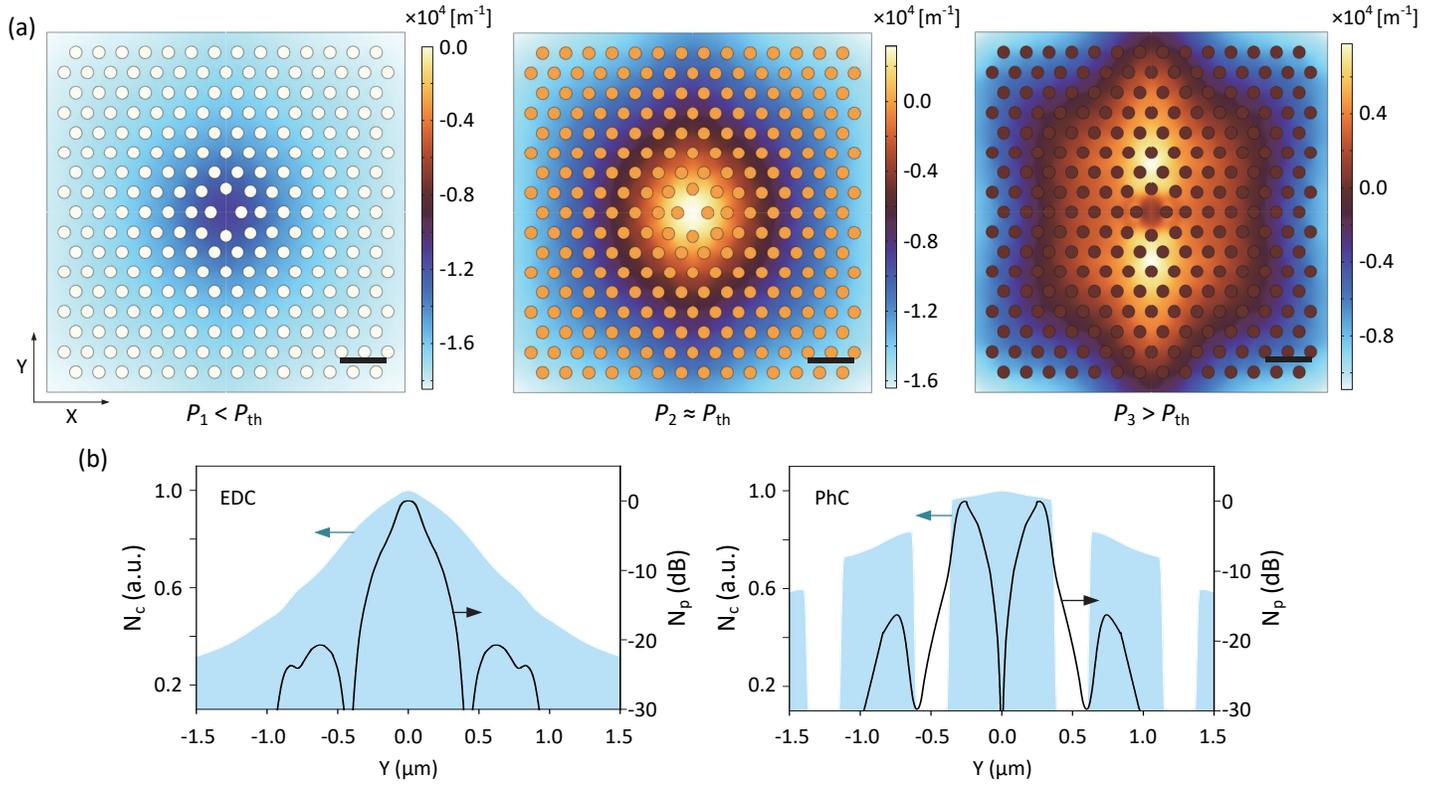

Fig. S4. (a) Calculated modal gain of the PhC H0 nanocavity laser at different power levels under 980 nm pumping: below (left), near (middle), and above (right) threshold. Scale bar: 1 μm. (b) Simulated photon (black curves) and carrier (blue shadings) density distributions at threshold under 980 nm pumping for the EDC (left) and PhC (right) lasers along the *y*-axis across the device centre. Values are normalized to their respective maxima.

Fig. S4 illustrates examples of the calculated modal gain and carrier density in the PhC laser. As seen, the modal gain is higher at the centre below and at threshold, but lower above threshold, which is attributed to spatial hole burning [20]. Simulations show that a higher spontaneous emission factor at the centre relative to the outer region (see Fig. S3) boosts the central concentration of carriers.

The underlying mechanism involves the smaller mode volume of the EDC cavity, which reduces the likelihood of encountering competing mode with high field intensity within the "hotspot". This configuration funnels carriers directly into the lasing mode instead of losing them to non-lasing modes, thereby enhancing laser efficiency. Consequently, while carrier consumption is intensified at the center due to spatial hole burning above laser threshold, fewer carriers are consumed below to at threshold where the field intensity is weak. We find this promotes a higher central carrier concentration even without localized central pumping.

**D. Interaction volume, laser threshold, and carrier volume**

We define the interaction volume $V_\mathrm{I}$ as a key metric for quantifying light-matter interactions:

$$V_\mathrm{I} = \frac{\left(\int N_c(\mathbf{r})d\mathbf{r}\right)\left(\int N_p(\mathbf{r})d\mathbf{r}\right)}{\int N_c(\mathbf{r})N_p(\mathbf{r})d\mathbf{r}} = \frac{1}{\int N_c(\mathbf{r})N_p(\mathbf{r})d\mathbf{r}}. \tag{S.13}$$

By employing the relations $N_p(\mathbf{r}) = \varepsilon(\mathbf{r})|E(\mathbf{r})|^2$ and $\int N_p(\mathbf{r})d\mathbf{r} = 1$, and performing a Taylor expansion of the gain term, we get from Eq. (S.12) that at threshold

$$\gamma_c = v_g g_N \int \left(N(\mathbf{r}) - \tilde{N}_{tr}(\mathbf{r})\right) N_p(\mathbf{r})d\mathbf{r}, \tag{S.14}$$

where $g_N = g_0/(N_{tr} + N_s)$ is the laser differential gain, and

$$\tilde{N}_{tr}(\mathbf{r}) = N_{tr} + \frac{1}{2}\frac{(N(\mathbf{r}) - N_{tr})^2}{N_{tr} + N_s} - \frac{1}{3}\frac{(N(\mathbf{r}) - N_{tr})^3}{(N_{tr} + N_s)^2} + \dots. \tag{S.15}$$

Then, by applying the relationship $\int N(\mathbf{r})d\mathbf{r} = \int n_c N_c(\mathbf{r})d\mathbf{r} = n_c$, with $\int N_c(\mathbf{r})d\mathbf{r} = 1$ and $n_c$ being the carrier number, and assuming $\tilde{N}_{tr}(\mathbf{r}) \approx n_{tr} N_c(\mathbf{r})$ so that $n_{tr}/V_\mathrm{I} = \int \tilde{N}_{tr}(\mathbf{r}) N_p(\mathbf{r})d\mathbf{r}$, where $n_{tr}$ is the total carrier number that must be inverted to reach transparency. The threshold carrier number is thus obtained as

$$n_{c,th} = V_\mathrm{I}\frac{\omega_c}{gQ} + n_{tr}, \tag{S.16}$$

which is proportional to $V_I$. In Eq. (S.16), $v_g g_N$ is replaced by $g$ for a more compact form. If taking the leading order in Eq. (S.15) (linearizing the gain term at threshold point), $n_{c,th}$ becomes $n_{c,th} = V_I \left( \dfrac{\omega_c}{gQ} + \Gamma N_{tr} \right)$. This form may suggest that lowering $n_{c,th}$ requires a smaller optical confinement factor $\Gamma$. However, as we will show, $V_I$ is inversely proportional to $\Gamma$, so a larger $\Gamma$ is still preferable.

It needs to be noted that the carrier number required to reach threshold must be smaller than the available states within the homogeneous bandwidth [21] and thus depends on the electron density of states of the active material $\rho(E)$, i.e., one should fulfil the relation

$$n_{c,a} = \min\{V_a, V_{pump}\} \times \left( \int \rho(E) f(E) dE \right) > n_{c,th}, \tag{S.17}$$

so that the needed gain can be provided. Here, $n_{c,a}$ is the maximum carrier number available, $f(E)$ is the Fermi-Dirac distribution, capped by the maximum pump strength realistically achievable, $\min\{V_a, V_{pump}\}$ represents the smaller value of the active region volume $V_a$ and the pumping volume $V_{pump}$. Satisfying Eq. (S.17) becomes challenging when $V_a$ is small and $\rho(E)$ transits from a continuous to a more discretized distribution with respect to energy ($E$). This situation arises when lateral quantum confinement is introduced, e.g., with an ultrasmall buried heterostructure [22] or quantum dots are used as the active material.

We now relate $V_I$ with the laser threshold power. As for the carrier rate equation, integrating both sides of Eq. (S.6) leads to the following equation at threshold

$$0 \approx \int D \nabla^2 N(\mathbf{r}) d\mathbf{r} - \int R_{sp}(N(\mathbf{r})) d\mathbf{r} - \int R_{nr}(N(\mathbf{r})) d\mathbf{r} + \int G_p(\mathbf{r}) d\mathbf{r}. \tag{S.18}$$

Here, we assume the net carrier flow outside the device is negligible, i.e., $\int D \nabla^2 N(\mathbf{r}) d\mathbf{r} \to 0$. On the other hand, one can replace it with an effective carrier loss, which can be absorbed into the nonradiative recombination term. By doing so, we arrive at

$$\int R_{sp}(N(\mathbf{r})) d\mathbf{r} + \int R_{nr}(N(\mathbf{r})) d\mathbf{r} \approx \int G_p(\mathbf{r}) d\mathbf{r} = \gamma_{th}, \tag{S.19}$$

where $\gamma_{th} = \int G_p(\mathbf{r}) d\mathbf{r} = P_{th}/(\hbar \omega_c)$ is the laser threshold pump rate. Since $R_{sp}(N(\mathbf{r})), R_{nr}(N(\mathbf{r})) \propto N(\mathbf{r})$, with $N(\mathbf{r}) \propto n_{c,th}$, the laser threshold power $P_{th}$ is proportional to $n_{c,th}$, and consequently, to $V_I$, i.e.,

$$P_{th} \propto V_I \frac{\omega_c}{gQ} + n_{tr}. \tag{S.20}$$

As seen, a smaller $V_I$ reduces the number of carriers needed to offset laser losses, thereby lowering the threshold power. As we can also find from Eq. (S.10), $R_{sp}$ is proportional to $1/\beta_i$. Since $\beta_i(\mathbf{r})$ is larger at the "hotspot" (see Fig. S3), a carrier distribution concentrated at the field maximum enhances the spatially-averaged spontaneous emission factor, further reducing $P_{th}$. In the following, we analyze the interaction volume $V_I$ in different regimes.

### D.1. Case 1: Carrier distribution varies significantly slower than photon distribution

If the carrier density varies slowly in space where the photon distribution peaks, $N(\mathbf{r})$ can be considered constant in the volume integral of Eq. (S.13). Consequently, $N(\mathbf{r}) = n_c N_c(\mathbf{r}) \approx n_c N_c$. When considering a finite lateral dimension of the active region, as in a laser with a buried heterostructure, we obtain

$$\int N(\mathbf{r}) d\mathbf{r} = n_c N_c \int_{\text{active}} d\mathbf{r} = n_c N_c V_a = n_c \Rightarrow N_c = \frac{1}{V_a}, \tag{S.21}$$

where $V_a$ is the volume of the active region. In this case, we find

$$\frac{1}{V_I} = \int N_c(\mathbf{r}) N_p(\mathbf{r}) d\mathbf{r} = N_c \int_{\text{active}} N_p(\mathbf{r}) d\mathbf{r} = N_c \frac{\int_{\text{active}} N_p(\mathbf{r}) d\mathbf{r}}{\int N_p(\mathbf{r}) d\mathbf{r}} = N_c \Gamma = \frac{1}{V_a/\Gamma} = \frac{1}{V_{op}}. \tag{S.22}$$

Eq. (S.22) shows that $V_I$ reduces to the conventional optical volume $V_{op} = V_a/\Gamma$ used in modelling traditional light sources [11]. Therefore, if neglecting the spontaneous emission, the threshold pump rate (Eq. (S.19)) becomes

$$\gamma_{th} \approx \gamma_{nr} V_a \left( \gamma_c / (\Gamma v_g g_N) + N_{tr} \right), \tag{S.23}$$

and the threshold power, $P_{th} = \hbar \omega_c \gamma_{nr} V_a \left( \gamma_c / (\Gamma v_g g_N) + N_{tr} \right)$, recovers to the form of macroscopic lasers.

### D.2. Case 2: Photon distribution varies significantly slower than carrier distribution

In the opposite case, where a single quantum emitter or a very small region of active material interacts with a slowly varying cavity mode in space, $N_p(\mathbf{r})$ can be assumed uniform across the integral in Eq. (S.13) so that

$$\frac{1}{V_\mathrm{I}} = \int N_c(\mathbf{r}) N_p(\mathbf{r}) d\mathbf{r} = N_p(\mathbf{r}_0), \tag{S.24}$$

where $\mathbf{r}_0$ is the spatial position of the quantum emitter. At the same time, we have

$$V_\mathrm{mod} = \frac{\int N_p(\mathbf{r}) d\mathbf{r}}{N_p(\mathbf{r}_0)} = \frac{1}{N_p(\mathbf{r}_0)} \Rightarrow \frac{1}{V_\mathrm{mod}} = N_p(\mathbf{r}_0). \tag{S.25}$$

Comparing Eqs. (S.24) and (S.25) reveals that $V_\mathrm{I}$ becomes the mode volume $V_\mathrm{mod}$ commonly used for quantum light sources. For a quantum emitter with a single electron in the excited state ($n_c = 1$), the spontaneous emission rate becomes

$$\int R_{sp}(N(\mathbf{r})) d\mathbf{r} = \gamma_{sp,i}(\mathbf{r}_0) + \sum_{n=1;n\neq i}^{\infty} \gamma_{sp,n}(\mathbf{r}_0).$$

In this situation, we arrive at the pumping rate required for driving single-photon sources, which operate in the spontaneous emission regime:

$$\gamma_{th} = \sum_{n=1}^{\infty} \gamma_{sp,n}(\mathbf{r}_0) + \gamma_{nr}. \tag{S.26}$$

**D.3. General situation**

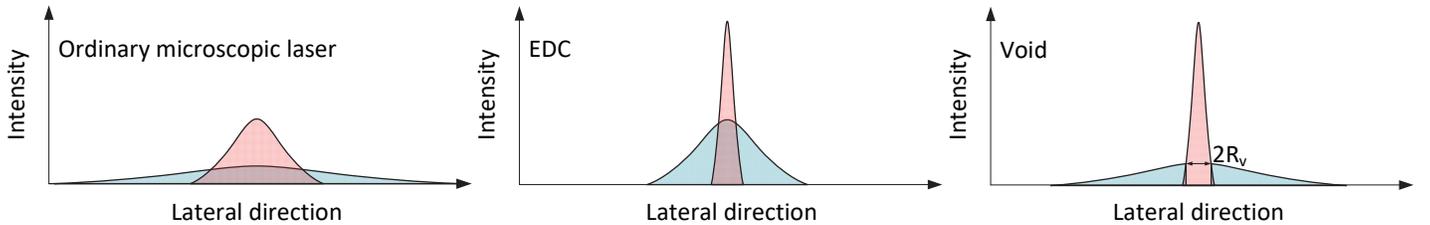

Fig. S5. Schematics illustrating the spatial distributions of photons (red) and carriers (blue) with Gaussian distributions and overlapping maxima for a conventional microscopic laser without carrier confinement (left), EDC laser (middle), and nanolaser with void region (right).

In general, $V_\text{I}$ depends on both $V_\text{mod}$ and the carrier volume $V_\text{car}$, which we define as $V_\text{car} = \int N(\mathbf{r})d\mathbf{r} / \max\{N(\mathbf{r})\}$. We consider a simple scenario where both the photon $N_p(\mathbf{r})$ and carrier $N_c(\mathbf{r})$ density have Gaussian distributions in the lateral (x-y) plane but are uniform along the z direction (Fig. S5). In polar coordinates,

$$N_p(\mathbf{r}) = N_{p0}\exp\left(-r^2/\left(2\sigma_p^2\right)\right), \quad N_c(\mathbf{r}) = N_{c0}\exp\left(-r^2/\left(2\sigma_c^2\right)\right)\Theta\left(r^2 - R_v^2\right), \quad \text{where} \quad r^2 = x^2 + y^2 \quad \text{and}$$

$\Theta\left(r^2 - R_v^2\right)$ represents a step function accounting for the absence of active material in void regions, as relevant for structures where photon distribution has its maximum in air gaps [6], [7], [23]. This void region is assumed to have a disc area of a radius $R_v$ at the centre (see Fig. S5), so $R_v=0$ corresponds to cases without the void region. $W_{p,c} = 2\sqrt{2\ln 2}\,\sigma_{p,c}$ is the full-width-at-half-maximum (FWHM) of the respective Gaussian profiles (both have their maxima at the centre).

These Gaussian functions have been normalized such that $\int N_{p,c}(\mathbf{r})d\mathbf{r} = 1$. In this case, one can deduce that $N_{p0} = 1/\left(2\pi h\sigma_p^2\right)$ and $N_{c0} = 1/\left(2\pi h_\text{QW}\exp\left(-R_v^2/\left(2\sigma_c^2\right)\right)\sigma_c^2\right)$, where $h$ is the membrane thickness. The mode and carrier volume are obtained as $V_\text{mod} = 1/N_{p0} = 2\pi h\sigma_p^2$ and $V_\text{car} = 1/\left(N_{c0}\exp\left(-R_v^2/\left(2\sigma_c^2\right)\right)\right) = 2\pi h_\text{QW}\sigma_c^2$, respectively. $V_\text{I}$ can thus be derived as

$$V_\text{I} = 1/\left(\int N_c(\mathbf{r})N_p(\mathbf{r})d\mathbf{r}\right) = (V_\text{mod} + V_\text{car}/\Gamma_z)\exp\left(R_v^2/\left(2\sigma_p^2\right)\right). \tag{S.27}$$

Here $\Gamma_z \approx h_\text{QW}/h$ is the vertical optical confinement factor. Simultaneously, the in-plane confinement factor becomes

$$\Gamma_{xy} = \frac{\int_0^{2\pi}\int_0^\infty N_p(r,\theta)rdrd\theta - \int_0^{2\pi}\int_0^{R_v} N_p(r,\theta)rdrd\theta}{\int_0^{2\pi}\int_0^\infty N_p(r,\theta)rdrd\theta} = \exp\left(-\frac{R_v^2}{2\sigma_p^2}\right). \tag{S.28}$$

Using Eq. (S.28) in Eq. (S.27) leads to

$$V_{\mathbf{I}} = \frac{V_{\text{mod}} + V_{\text{car}}/\Gamma_z}{\Gamma_{xy}} = \frac{V_{\text{mod}}}{\Gamma_{xy}} + \frac{V_{\text{car}}}{\Gamma}, \tag{S.29}$$

where $\Gamma = \Gamma_{xy}\Gamma_z$ is the total optical confinement factor. Eq. (S.28) shows that $\Gamma_{xy}$ in a void/slot structure where $R_v \geq 0.5 W_c$ is no larger than 0.5, consistent with the results in Ref. [24]. In contrast, in non-void cases, one typically has $\Gamma_{xy} \to 1$. Then, the interaction volume is reduced to

$$V_{\mathbf{I}} = (V_{\text{mod}} + V_{\text{car}}/\Gamma_z)/\Gamma_{xy} \approx V_{\text{mod}} + V_{\text{car}}/\Gamma. \tag{S.30}$$

As revealed, $V_{\text{car}}/\Gamma$ plays a role similarly to the optical volume $V_{op}$ for macroscopic lasers [11]. Eq. (S.30) illustrates that when $V_{\text{mod}} \ll V_{\text{car}}$ ($V_{\text{mod}} \gg V_{\text{car}}$), $V_{\mathbf{I}}$ aligns with the scenarios described in section D.1 (D.2).

**E. Additional data**

**E.1. Pump-excited field patterns**

We analyze the field patterns excited by the pump beam using FDTD simulations. A pump laser beam at 1310 (980) nm with a beam diameter of ~2.5 (2.2) mm passes through an objective lens with a numerical aperture of 0.65, a lens diameter of 5.2 mm, and a focal length of approximately 4 mm. This setup produces a Gaussian beam with a $1/e^2$ spot diameter of ~2.5 (2) μm at the focal plane, which agrees very well with our measured values. The simulations encompass the vectorial nature of the Gaussian beam and the entire device structure, including the InP membrane and Si substrate, with the pump light polarized along the $y$-direction to maximize efficiency. Despite the initial Gaussian profile, the excited field ($|\mathbf{E}_p|$) across the QW plane exhibits a more complex pattern, cf. Figs. S6-S8. This pattern ($|\mathbf{E}_p|^2$, due to the dominant linear absorption) is subsequently imported as $G_p(x,y)$ into Eq. (S.6) for laser simulations.

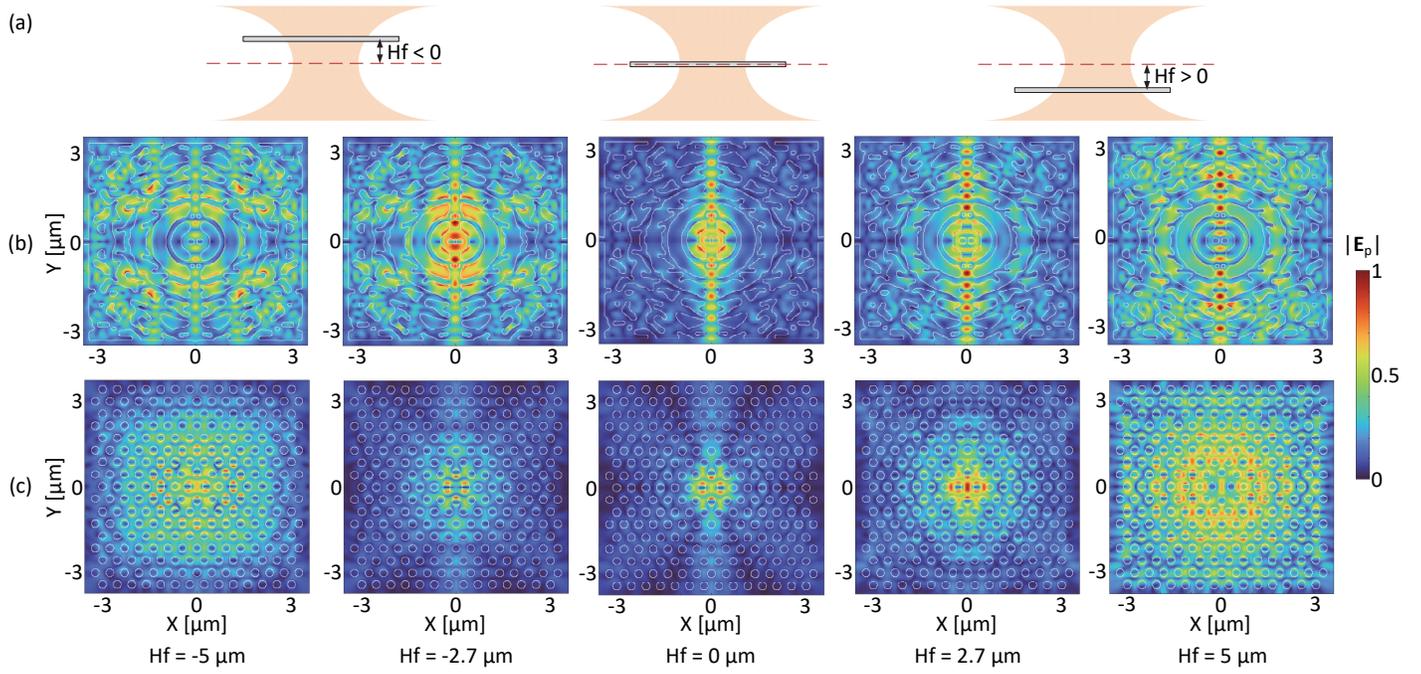

Fig. S6. (a) Schematic of the excitation, where a 1310 nm Gaussian beam (orange shading) from an objective lens strikes the membrane (gray slab) containing the EDC or PhC nanocavity. Dashed red lines indicate the focal plane of the Gaussian beam. The left (right) plot shows a negative (positive) focal height, Hf, with the focus below (above) the membrane, and the middle plot represents Hf=0. (b) Simulated pump excitation patterns ($|E_p|$, normalized) at the central plane of the EDC membrane. (c) Same as (b) but for the PhC nanocavity. The plots correspond to Hf ranging from -5 μm to 5 μm.

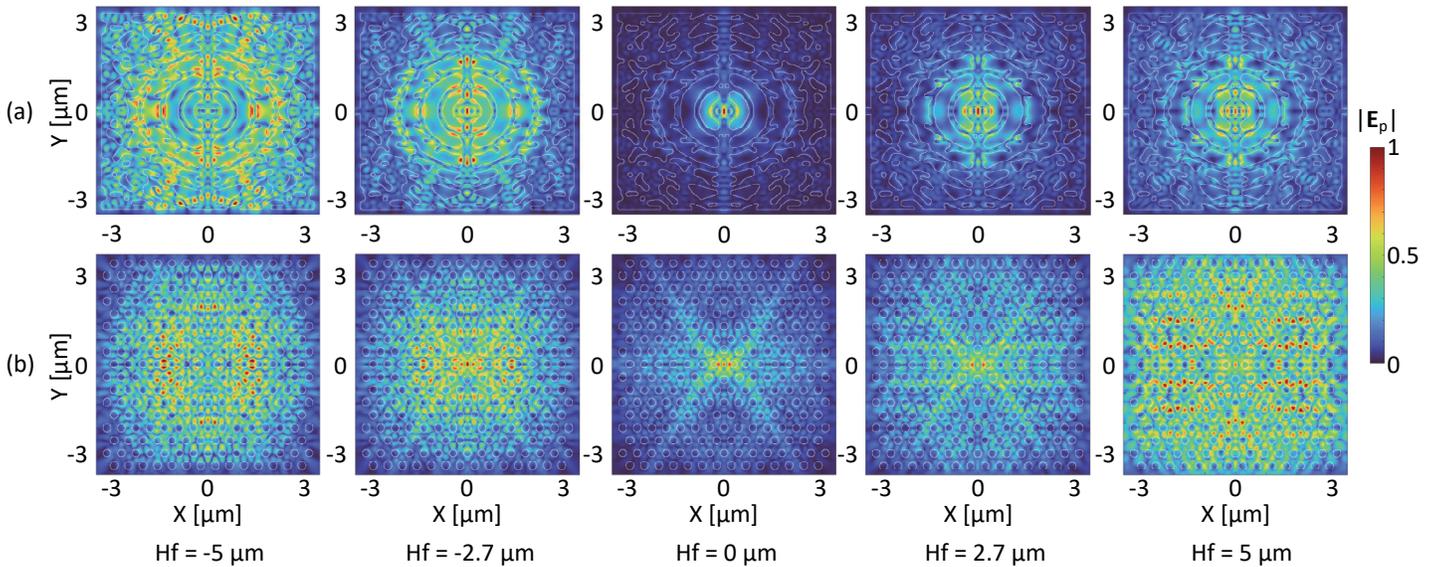

Fig. S7. Same as Fig. S6 but with 980 nm pump light. Excitation patterns for the (a) EDC and (b) PhC nanocavity.

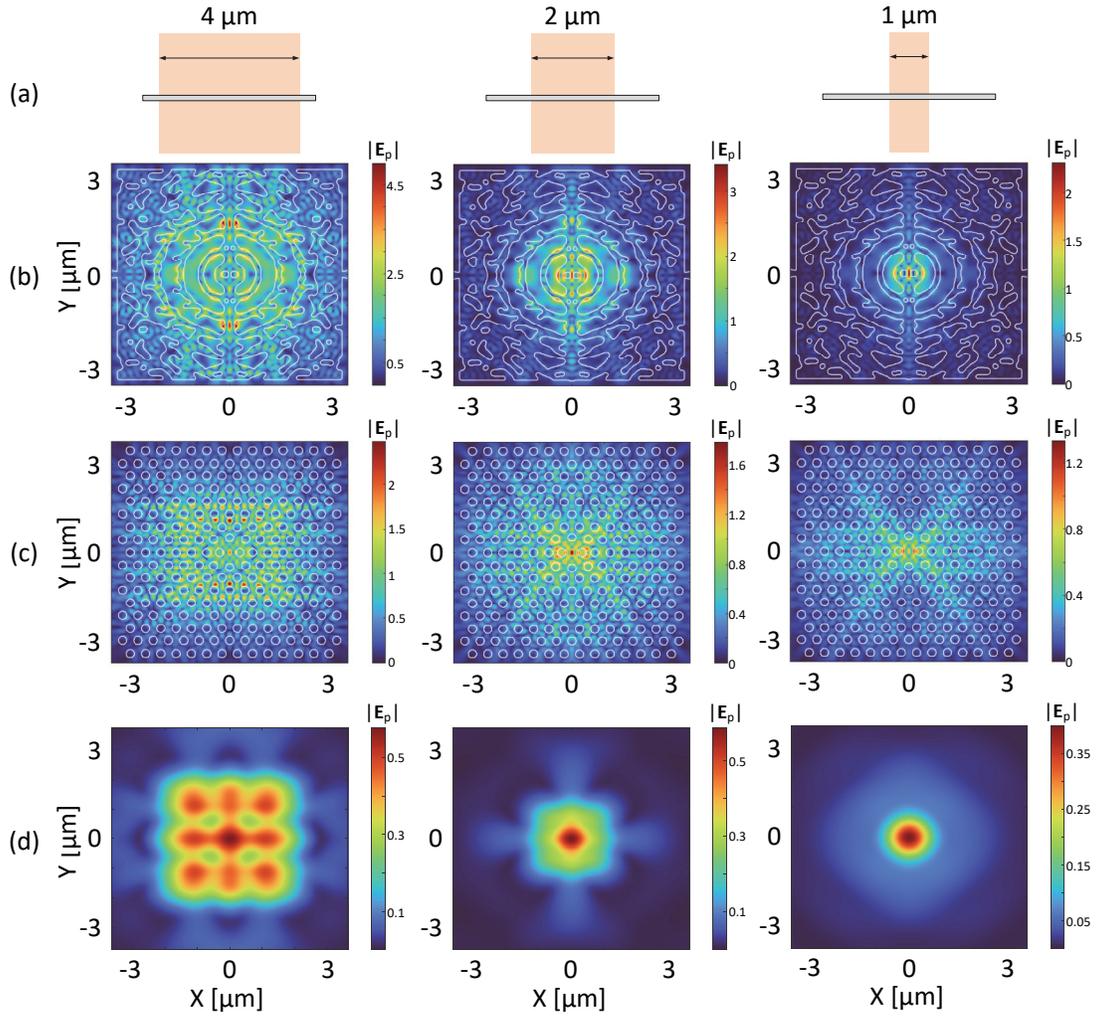

Fig. S8. (a) Similar to Fig. S7 but with plane-wave excitation (orange shading) from a square window of varying side lengths. The pump light before striking the device uniformly maintains an intensity of $|\mathbf{E}_p|=1$ across the window. (b) Simulated pump excitation patterns at the central plane of the EDC membrane ($|\mathbf{E}_p|$ not normalized). (c) Same as (b) but for the PhC membrane. (d) Same as (b) and (c) but for a non-structured membrane. From left to right, the side lengths of the pump square window are 4 µm, 2 µm, and 1 µm.

Here, we define the vertical distance between the focal plane of the pump beam and the device's central plane as |Hf|, where positive (negative) Hf means the pump focus is above (below) the membrane, causing a diverging (converging) beam cross the membrane. Fig. S7 shows that at Hf=0, the excited pattern at 980 nm closely resembles the lasing mode, with one antinode at the centre for the EDC and two antinodes near the central holes for the PhC H0 nanocavity.

Fig. S6 shows that the EDC laser exhibits a more dispersed excitation pattern at 1310 nm than the PhC laser, meaning a less localized carrier generation profile. However, the EDC still achieves a lower threshold, demonstrating its superior ability to slow carrier diffusion. Fig. S8 shows the pump pattern from plane-wave injection at varying spot sizes. As the pump spot size decreases, fewer cavity modes are excited, rendering the pattern less homogeneous and increasing the selective excitation of a mode pattern and, subsequently, a carrier distribution that aligns closer with the lasing mode.

**E.2. Additional measurement data for EDC and PhC lasers**

Figure S9 includes the 980 nm CW pumping measurements and additional measurements of the PhC laser.

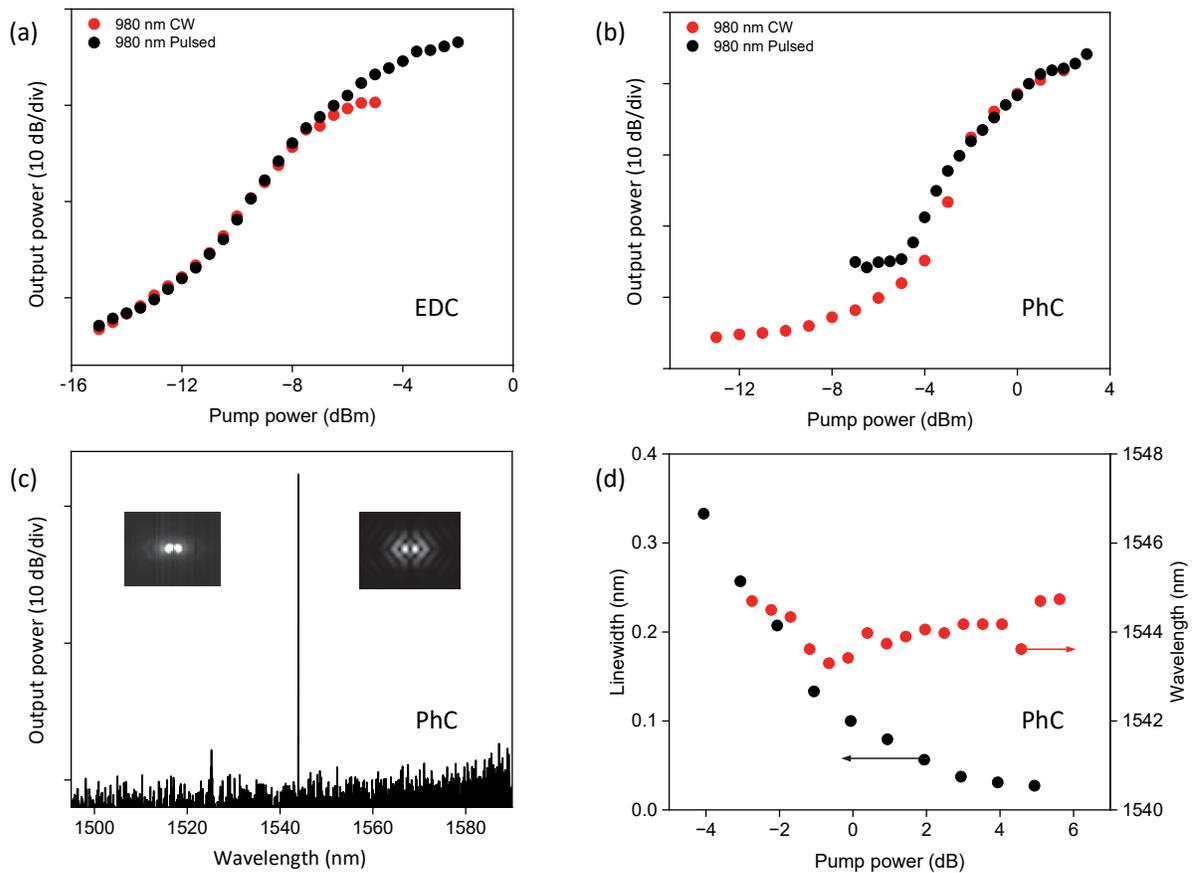

Fig. S9. (a, b) Input-output curves of the EDC (a) and PhC (b) laser under 980 nm CW (red) and pulsed (black) pumping. (c) Lasing spectrum of the PhC H0 laser above threshold. Inset: measured (left) and simulated (right) emission patterns. (d) Measured linewidth (black) and lasing wavelength (red) versus pump power for the PhC H0 laser. The pump power is normalized with zero value at threshold. The linewidth is limited by the resolution (0.02 nm) of our optical spectrum analyzer at high pump powers. In (c) and (d), the pump is the 1310 nm CW light source, as Fig. 2. All measurements are performed at room-temperature.

Figure S9(b) shows that the PhC laser's threshold curve remains largely unchanged between pulsed and CW pumping (the output power "flattening" at low input power under pulsed pumping is due to noise floor saturation on our spectrum analyzer, resulting from a lower signal-to-noise ratio compared to CW pumping). In contrast, the EDC laser displays a more pronounced S-curve under pulsed pumping (Fig. S9(a)), similar to Fig. 3a, indicating stronger thermal effects due to enhanced carrier and field localization. This pronounced S-curve better aligns with simulation results. The larger thermal effect is further supported by the larger lasing wavelength redshift in the EDC laser (Fig. 2c) compared to the PhC laser (Fig. S9(d)). While surface passivation aids in mitigating the thermal effects, local heating cannot be completely eliminated. Further improvement could be made by encapsulating the device in materials with higher thermal conductivity [25].

The highest collected power from the EDC (PhC) laser is around -40 dBm (-48 dBm), with their collection efficiency ratio closely matching simulations. This difference is mainly due to the EDC laser's lower vertical (intrinsic) $Q$-factor and more centralized emission. The EDC laser features a single antinode at the centre (insets in Fig. 2a), in contrast to the dual antinodes of the PhC H0 laser (insets in Fig. S9(c)).

Figures S10 and S11 illustrate the threshold characteristics as the focal height ($H_f$) varies. It is important to note that both the pumping and collection efficiency decrease with larger $|H_f|$ (e.g., $|H_f| = 5$ μm), causing a drop in signal-to-noise ratio and reducing the power range with fewer reliable data points. To ensure a more transparent comparison between the lasers, the input and output powers have been normalized, keeping the QW absorbed power (or electrical energy in the QW area) the same for both types, as determined by

$$\int_{\text{EDC},\varepsilon>\varepsilon_0} \varepsilon(x,y)|\mathbf{E}_p(x,y)|^2 dxdy = \int_{\text{PhC},\varepsilon>\varepsilon_0} \varepsilon(x,y)|\mathbf{E}_p(x,y)|^2 dxdy. \tag{S.31}$$

Here, the labels EDC and PhC beneath the integral sign specify that the integration is being performed over the central plane of the respective structures.

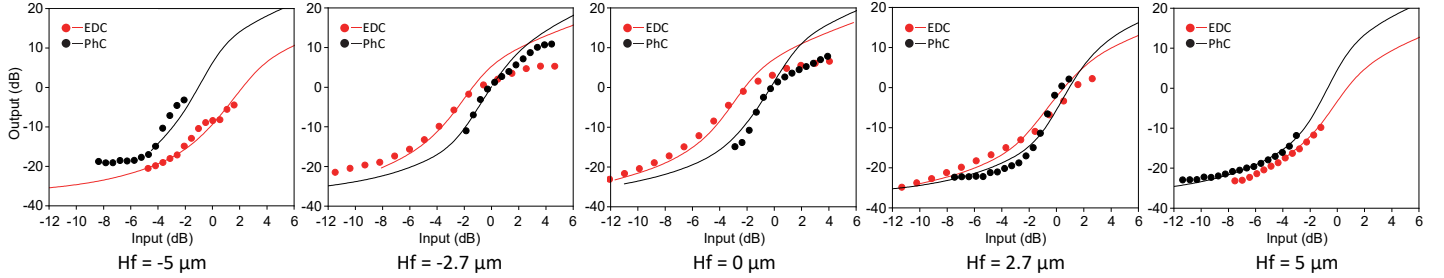

Fig. S10. Threshold characteristics of the EDC (red) and PhC (black) nanolasers for focal heights (Hf) ranging from -5 μm to 5 μm, from left to right. The CW pump at 1310 nm is used. Dots represent experimental data and solid curves show numerical simulations of the 2D laser model. The input power is normalized according to Eq. (S.31), ensuring that the absorbed power is identical for both laser types. Subsequently, both input and output powers are shifted on the log-log scale for clearer comparison across different cases of Hf, while the relative input power between the EDC and PhC lasers is maintained.

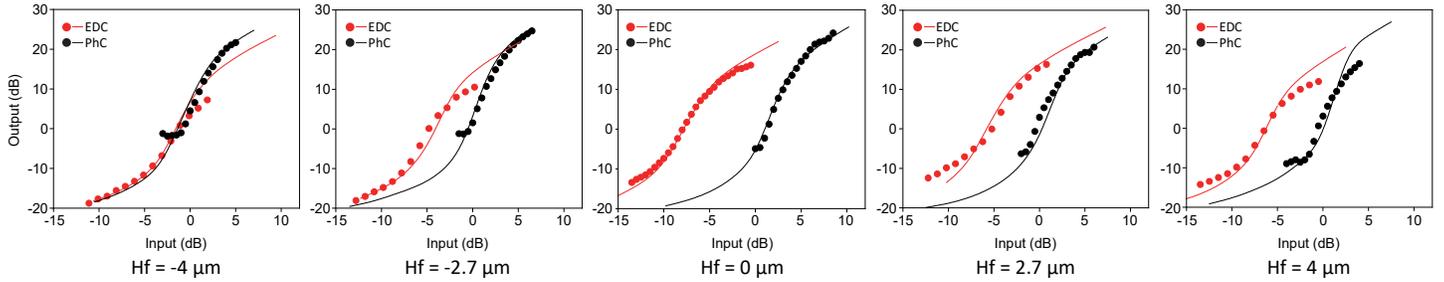

Fig. S11. Same as Fig. S10 but with the pulsed pump at 980 nm.

The laser threshold power is determined as the peak of the first derivative of output power with respect to input, calculated on a log-log scale from theoretical fits. This method has proven effective across a wide range of cases [22], [26]. Twenty nominally identical devices for each laser type are measured across the wafer. Fig. S12 displays the measurements by varying the focal plane of the pump light, averaged across these 20 devices. Following Eq. (S.31), the threshold values are normalized, with the PhC H0 laser's lowest point set to zero. As seen, both lasers reach a threshold minimum at the focal plane and increase as they move away. Although the threshold reduction for the EDC laser appears smaller at 980 nm before pump power normalization (comparing Figs. 3a and 3b), the reduction is actually larger after the normalization (Fig. 4c).

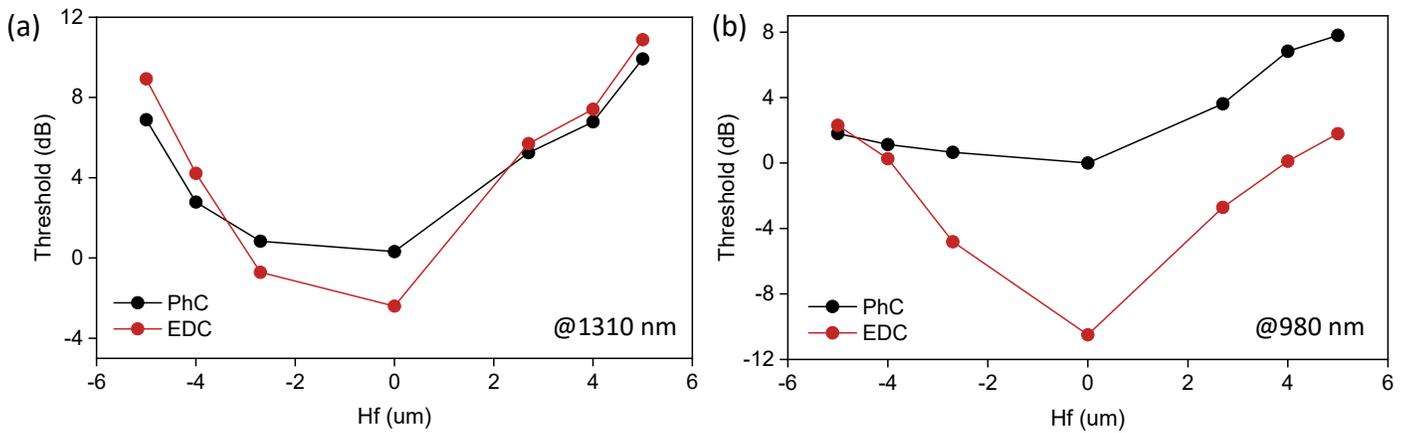

Fig. S12. Measured variation in the thresholds of the EDC (red) and the PhC (black) lasers (averaged over 20 devices) for different focal plane positions of the pump light. (a) and (b) correspond to pumping at 1310 nm and 980 nm, respectively. The focal height, $H_f$, ranges from -5 μm to 5 μm.

**E.3. EDC laser with a narrower dielectric bridge**

An advantage of the EDC cavity is its capability to further reduce $V_{mod}$ well below the diffraction limit by simply reducing the feature size of the central dielectric nanostructure [9], [27], as exemplified in Fig. S13(a). However, in our case, reducing the bridge width below 80 nm results in only a gradual decrease in $V_{mod}$ but a sudden drop in the in-plane optical confinement factor $\Gamma_{xy}$ due to the mode field expansion into the air. Even lower $V_{mod}$ is attainable with tipped or bowtie geometries [9], [27] instead of the bridge form. However, according to our simulations, these geometries are found to accelerate carrier diffusion, increasing $V_{car}$ (and $V_I$) and result in a higher laser threshold.

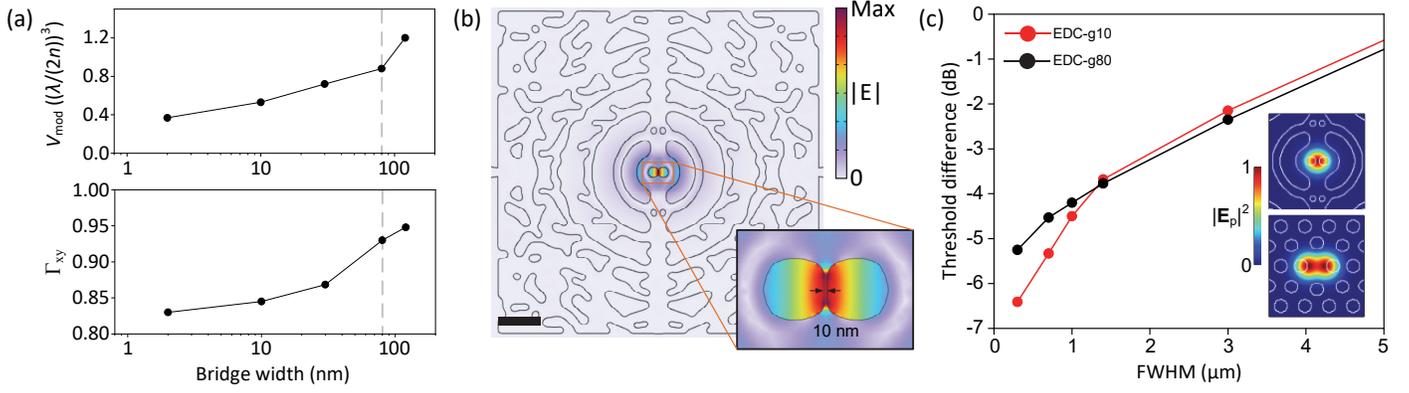

Fig. S13. (a) Calculated mode volume $V_{mod}$ (upper) and in-plane optical confinement factor $\Gamma_{xy}$ (lower) versus the nanobridge width in our EDC cavity. The dashed lines mark the structure used. (b) Lasing mode ($|\mathbf{E}|$) in the modified EDC cavity (EDC-g10) analogous to the structure (EDC-g80) in Fig. 1, but with a reduced central bridge width of 10 nm. Scale bar: 1 μm. The EDC-g10 is assumed to retain the same resonant wavelength and $Q$-factor as EDC-g80. Inset: zoom-in of the central mode field. (c) Simulated variation in the threshold power difference of the EDC-g10 (red) and EDC-g80 (black) relative to the PhC laser for varying FWHM of the Gaussian excitation pattern ($|\mathbf{E}_p|^2$). The pump power has been normalized according to Eq. (S.31). Insets show the Gaussian excitation patterns for the EDC (upper) and PhC (lower) lasers. The excitation pattern for the EDC lasers is a single Gaussian centred on the EDC mode's "hotspot", while it is a dual-Gaussian peaking at the two antinodes of the PhC H0 nanocavity's fundamental mode.

To explore the effect of a smaller $V_{mod}$, we modified the existing EDC structure (EDC-g80) by reducing the bridge width from 80 nm to 10 nm (EDC-g10) via bringing the two central neighbouring air holes closer, cf. Fig. S13(b). The 5 nm thick $Al_2O_3$ layer (neglected here), in fact, limits the bridge width to ~10 nm, approaching our current fabrication limit [9]. This adjustment results in a $V_{mod}$ of $0.53(\lambda/2n)^3$. Typically, reducing $V_{mod}$ also lowers the $Q$-factor, e.g., the 10 nm bridge decreases the $Q$-factor to ~900 and shifts the wavelength to ~1410 nm. A slimmer bridge also lowers thermal conductivity, heightens perturbation sensitivity, and may intensify gain saturation due to lateral quantum confinement. Here, we neglect all these effects and assume unchanged resonant frequency and $Q$-factor as our EDC-g80 (potentially through a new round of topology optimization). We focus solely on the impact of a reduced $V_{mod}$ and simplify the excitation pattern as Gaussian functions with varying FWHM.

The EDC laser consistently shows a lower threshold than the PhC laser, especially as the excitation pattern size diminishes (Fig. S13(c)). The EDC-g10 exhibits a higher threshold than EDC-g80 under large pump sizes, likely due to a smaller $\Gamma_{xy}$. The advantages of EDC-g10 over EDC-g80 become evident primarily under ultrasmall localized pumping, underscoring the more important roles of $V_{car}$ and $V_I$ over $V_{mod}$ to boost light-matter interactions in structures with minimal $V_{mod}$. Future improvements in EDC laser may include local current injection or other advanced pumping schemes [28], high-density quantum dots with minimized inhomogeneous broadening [29], or buried heterogeneous structures [22] with engineered dimensions. We also performed simulations on structures similar to Refs. [6], [23] where the field is maximized in non-active regions (not shown here), which exhibit higher thresholds than the PhC laser, consistent with findings in Ref. [7]. Further analysis will be detailed in future publications.